\documentclass[aps,prfluids,onecolumn,superscriptaddress]{revtex4-2}

\usepackage{graphicx}
\usepackage{amsmath}
\usepackage{amssymb}
\usepackage{xcolor}
\usepackage{tikz}
\usetikzlibrary{shapes}
\usepackage{pgf} 
\usepackage{siunitx}
\usepackage{booktabs} 
\usepackage{multirow} 
\usepackage{bigdelim}
\usepackage{amssymb,pifont}
\usepackage{soul}

\definecolor{color_ha}{rgb}{0.2422,0.1504,0.6603}
\definecolor{color_hb}{rgb}{0.2517,0.4261,0.9974}
\definecolor{color_hc}{rgb}{0.0628,0.6972,0.8355}
\definecolor{color_hd}{rgb}{0.4322,0.8028,0.4013}
\definecolor{color_he}{rgb}{0.9962,0.7798,0.2095}

\definecolor{color_haContour}{rgb}{0.2809,0.2875,0.9305}
\definecolor{color_hbContour}{rgb}{0.1540,0.5902,0.9218}
\definecolor{color_hcContour}{rgb}{0.1853,0.7721,0.6379}
\definecolor{color_hdContour}{rgb}{0.8281,0.7481,0.1536}
\definecolor{color_heContour}{rgb}{0.9692,0.9609,0.1061}

\definecolor{color_violet}{rgb}{0.62,0.34,0.64}
\definecolor{color_Blue}{rgb}{0, 0, 1}

\usepackage{subfigure}

\begin{document}

\newcommand{\markerhaSimu}{\raisebox{0.5pt}{\tikz{\node[draw,scale=0.5,regular polygon, regular polygon sides=3,fill=color_ha,draw=color_haContour](){};}}}
\newcommand{\markerhbSimu}{\raisebox{0.5pt}{\tikz{\node[draw,scale=0.5,regular polygon, regular polygon sides=3,fill=color_hb,draw=color_hbContour, rotate = 180](){};}}}
\newcommand{\markerhcSimu}{\raisebox{0.5pt}{\tikz{\node[draw,scale=0.6,regular polygon, regular polygon sides=4,fill=color_hc,draw=color_hcContour,rotate=45](){};}}}
\newcommand{\markerhdSimu}{\raisebox{0.5pt}{\tikz{\node[draw,scale=0.7,circle,fill=color_hd,draw=color_hdContour](){};}}}
\newcommand{\markerheSimu}{\raisebox{0.5pt}{\tikz{\node[draw,scale=0.7,regular polygon, regular polygon sides=4,,fill=color_he,draw=color_heContour](){};}}}

\newcommand{\markerha}{\raisebox{0.5pt}{\tikz{\node[draw,scale=0.5,regular polygon, regular polygon sides=3,fill=none,color=color_ha](){};}}}
\newcommand{\markerhb}{\raisebox{0.5pt}{\tikz{\node[draw,scale=0.5,regular polygon, regular polygon sides=3,fill=none,color=color_hb, rotate = 180](){};}}}
\newcommand{\markerhc}{\raisebox{0.5pt}{\tikz{\node[draw,scale=0.6,regular polygon, regular polygon sides=4,fill=none,color=color_hc,rotate=45](){};}}}
\newcommand{\markerhd}{\raisebox{0.5pt}{\tikz{\node[draw,scale=0.7,circle,fill=none,color=color_hd](){};}}}
\newcommand{\markerhe}{\raisebox{0.5pt}{\tikz{\node[draw,scale=0.7,regular polygon, regular polygon sides=4,,fill=none,color=color_he](){};}}}

\newcommand{\markerhD}{\raisebox{0.5pt}{\tikz{\node[draw,scale=0.5,regular polygon, regular polygon sides=3,fill=black,draw=black,rotate = 90](){};}}}
\newcommand{\markerLD}{\raisebox{0.5pt}{\tikz{\node[draw,scale=0.5,regular polygon, regular polygon sides=3,fill=red,draw=red,rotate = -90](){};}}}

\newcommand{\markerExpEvolution}{\raisebox{0.5pt}{\tikz{\node[draw,scale=0.5,circle,fill=none,color=color_Blue](){};}}}
\newcommand{\markerSimuTime}{\raisebox{0.5pt}{\tikz{\node[draw,scale=0.5,circle,fill=color_Blue,color=color_Blue](){};}}}


\title{Vortex ring induced by a disk translating toward or away from a wall 
~}


\author{Joanne Steiner}
\email[]{joannesteiner@hotmail.fr}
\affiliation{Universit\'e Paris-Saclay, CNRS, Laboratoire FAST, F-91405 Orsay, France}
\author{Cyprien Morize}
\email[]{cyprien.morize@universite-paris-saclay.fr}
\affiliation{Universit\'e Paris-Saclay, CNRS, Laboratoire FAST, F-91405 Orsay, France}

\author{Ivan Delbende}
\affiliation{Sorbonne Universit\'e, CNRS, Institut Jean Le Rond d'Alembert, 75005 Paris, France}

\author{Alban Sauret}
\affiliation{University of Maryland, College Park, Department of Mechanical Engineering, MD 20742, USA}
\affiliation{University of Maryland, College Park, Department of Chemical and Biomolecular Engineering, MD 20742, USA}

\author{Philippe Gondret}
\email[]{philippe.gondret@universite-paris-saclay.fr}
\affiliation{Universit\'e Paris-Saclay, CNRS, Laboratoire FAST, F-91405 Orsay, France}


\date{\today}

\begin{abstract}

This study investigates the time evolution of vortex rings generated by the normal translation of a disk either toward or away from a wall. We systematically vary the control parameters, including the disk size, stroke length, travel time, and distance from the wall, to analyze their influence on vortex dynamics. Experiments are conducted with Particle Image Velocimetry, while numerical simulations are performed using the flow solver Basilisk. The circulation and core radius are used as primary metrics to describe the properties of the vortex ring. A quantitative agreement is observed between the experimental and the numerical results. Both approaches reveal that the vortex circulation increases for disk motions in either direction — toward or away from the wall. However, the effect of the bottom wall on the vortex core radius differs depending on the motion of the disk. The presence of the wall increases the core radius only when the disk moves toward the wall, while no significant effect is observed for a translation away from the wall. Furthermore, we establish scaling laws to describe the maximum circulation and core size of the vortex ring as a function of the control parameters: the disk diameter, the typical time and stroke length of its motion and the minimal distance from the wall. These scalings, which differ from the unbounded case, contribute to a deeper understanding of vortex dynamics in the vicinity of solid boundaries.
\end{abstract}


\maketitle


\section{Introduction} \label{SecI}

The formation of vortices is ubiquitous in engineering and biological systems and, as a result, has attracted significant attention and has inspired numerous studies. In nature, many animals, such as insects, birds, and fish, generate arrays of vortices as they flap their wings, fins, or tails for locomotion \cite{Ellington_leadingEdge_1996, PullinWang2004, Wu_2011}. Dandelion seeds also generate vortices in their wake during flight \cite{cummins_separated_2018}. In engineering, the wingtip vortices generated in the wake of aircraft have been widely studied due to their impact on flight performance \cite{devenport_1996}. Similarly, vortex-induced vibration is a critical consideration in the building of civil engineering structures, such as bridges or offshore structures \cite{CHKWilliamson_2004}. Furthermore, wake vortices play also a role in energy harvesting systems such as wind turbines or flapping wing devices \cite{Whale2000, Xiao_2014}.

In 1904, \citet{Prandtl1904} considered theoretically the problem of vortex formation in the wake of translating objects after experimental observations such as the rolling up of a vortex sheet. Building on this work, \citet{Kaden1931} developed a self-similar theory for the winding of a vortex sheet. He used the potential flow associated with the translation of a plate at uniform velocity $U$ during the time $t$, which generates a free-rolling vortex sheet when the solid is instantaneously removed from the flow, and showed that the radius of the vortex should grow with time as $t^{2/3}$. Based on these results, \citet{wedemeyer_ausbildung_1961} predicted the time evolution of the circulation and the position of a vortex in the wake of a plate translating at a constant velocity. Since then, many studies have focused on the formation of such vortices, both experimentally \cite{Pierce1961, Pullin1980, Taneda1971, Steiner_2023} and numerically \cite{Xu_Nitsche_2014, Steiner_2023}. In particular, it has been shown that the size of the vortex is proportional to the plate length (or disk diameter) $D$ to the power of $1/3$ and to the stroke length $L$ to the power of $2/3$ \cite{Taneda1971, Steiner_2023} and that the circulation is proportional to $D^{2/3}$ and $L^{4/3}$ and inversely proportional to the travel time $\tau$ \cite{Steiner_2023}. For the flow past a circular disk, the generated vortex ring remains axisymmetric when the stroke length is not too large ($L/D < 4$) \cite{Yang2012} and in the case of a steady regime when the Reynolds number is  not too large \cite{Shenoy2008}.

While much of the early work focused on vortices in unbounded domains, some studies have explored how walls and boundaries influence vortex formation and dynamics. Indeed, as the body translates toward or away from a boundary, interactions with the wall modify the rolling up of the vortex sheet. \citet{walker_impact_1987} demonstrated experimentally such an interaction by describing the collision of a vortex ring with a wall. The formation of a significant unsteady boundary layer eventually leads to the detachment of a secondary vortex ring, causing the primary vortex to bounce off the wall \cite{walker_impact_1987, orlandi_vortex_1993, swearingen_dynamics_1995, pellerin_interaction_1999,naguib_2004}. Vortex interaction with the ground is responsible for the so-called ground effect: animals swimming or flying close to the ground experience benefits in steady-state configurations \cite{Rayner_1991,FernandezPrats2015}. In transient configurations, some studies have highlighted the benefits of the ground effect for insect takeoffs \cite{VanTruong_2013}, while others have not for swimming stingrays \cite{Blevins_2013}. The thrust produced by a rigid or flexible \cite{quinn_flexible_2014} pitching foil close to the ground is experimentally increased compared to the configuration far from the ground, also suggesting that the ground effect could help the motion. Moreover, the displacement of contaminants by footsteps is closely related to the formation of eddies and their interaction with a sediment bed \cite{kubota_experimental_2009, khalifa_particle_2007}. Finally, in some configurations, heave plates from offshore structures oscillate near the seabed, leading to a change in the added mass and damping coefficients \cite{garrido_2015}.

Although some fundamental mechanisms of vortex-wall interactions have been identified, a quantitative characterization (circulation, core size, trajectory) and modeling of vortices generated in the vicinity of a wall remain to be developed.
In the present work, such vortices are generated experimentally and numerically either by moving a disk toward a plane wall or away from the wall. The experimental configuration and the numerical methods are presented in Sec. \ref{SecII} together with the methodology used to characterize the main features of the vortex ring: its circulation, core radius, and trajectory. The results for a disk moving toward a wall are presented in Sec. \ref{SecIII}, while those for a disk moving away from the wall are presented in Sec. \ref{SecIV}. In both cases, the differences from the unbounded case are highlighted, and scaling laws accounting for the presence of the bottom wall are provided.

\section{Flow configuration and methods} \label{SecII}

\subsection{Flow configuration}

\begin{figure}[t]
     \centering
     \includegraphics[width=0.5\linewidth]{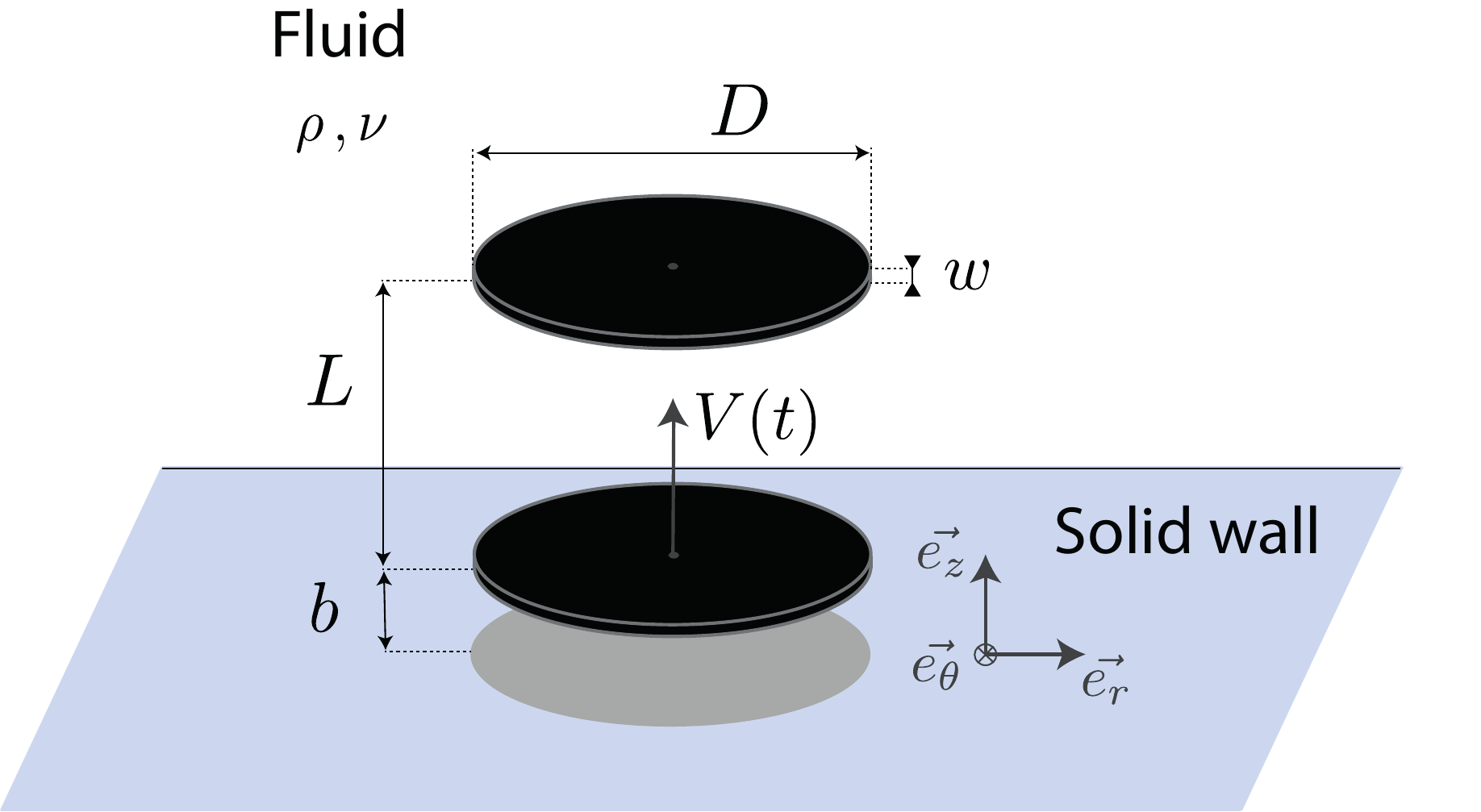}
        \caption{Sketch of the configuration with the notations used.}
        \label{fig:Fig1}
\end{figure}

We consider the configuration of a disk of diameter $D$ subjected to an unsteady translation over the stroke length $L$ during the time $\tau$ toward or away from a minimal distance $b$ of a parallel solid wall, as sketched in Fig. \ref{fig:Fig1}. The time evolution of the distance $h(t)$ of the disk from the wall is given by  
\begin{equation}
    h(t) = b+(L/2)[1 \pm \cos{(\pi t/\tau)}],
    \label{eq:DiskPosition}
\end{equation}
for $0 \leqslant t \leqslant \tau$, where the plus or minus sign refers to motion toward or away from the wall, respectively. If the disk approaches the wall, it starts at the initial time $t = 0$ at the distance $b + L$ from the wall and stops at $t = \tau$ at the minimum distance $b$. When the disk moves away from the wall, it starts at its minimum distance $b$ from the wall and stops at distance $b + L$ after a time $\tau$. 
The speed of the disk is thus 
\begin{equation}
\frac{dh}{dt} (0 \leqslant t^*\leqslant 1) = \mp V_m \sin(\pi t^*),
\label{eq:DiskVelocity}
\end{equation}
where $V_m = \pi L/2 \tau$ is the maximum velocity of the disk reached at mid-stroke and $t^* = t/\tau$ is the reduced dimensionless time. A positive (resp. negative) disk velocity corresponds to a disk moving away from (resp. toward) the wall. The sinusoidal motion has been chosen to allow a smooth evolution of the velocity with a finite acceleration and also as a first step toward the case of periodic oscillations. We will use the cylindrical coordinates $(r, \theta, z)$ with the axis $z$ corresponding to the disk axis and $z = 0$ at the minimum distance $b$ from the wall. 

The three relevant dimensionless quantities of the flow are the Reynolds number $Re = V_m D/\nu$, where $\nu$ is the kinematic viscosity of the fluid, the relative stroke length $L/D$, and the relative minimum distance between the disk and the wall $b/D$. We will restrict ourselves to an inertial regime of sufficiently high Reynolds number, above $10^3$, where the fluid viscosity does not significantly alter the characteristics of the vortices generated \cite{Steiner_2023}.

\subsection{Experimental methods}
The experiments were performed in a rectangular tank with a square cross-section of 40 cm $\times$ 40 cm and a height of 60 cm. The tank is filled with water of kinematic viscosity $\nu = 10^{-6}~$m$^2$/s at ambient temperature over a height of 40 cm. The disk consists of a thin acrylic plate of thickness $w$ = 2 mm and diameter 5 cm $\leqslant D \leqslant$ 15 cm such that $w/D \ll 1$. The disk is far from any side walls but close to the bottom wall with a distance varied between $0.2~$cm$~\leqslant~b~\leqslant~2~$cm ($b/D < 1$). A vertical rod of diameter 1 cm is attached to the center of the disk and connected to an eccentric system that transforms the rotation of a motor into a sinusoidal translation of the disk. The stroke length $L$ can be varied up to 5.2 cm by changing the position of the eccentric. The time $\tau = \pi/\Omega$, where $\Omega$ is the angular velocity of the motor, corresponds to the time the disk takes to travel the stroke length $L$ and was varied between 0.36 s and 2.5 s. The non-dimensional numbers were varied experimentally in the range $Re \in [1.3\times10^3, ~2\times10^4]$, $L/D \in [0.19,~0.56]$ and $b/D \in [0.01,~0.27]$ and are summarized in Table \ref{tab:Symbols}. As the dimensionless stroke length $L/D$ remains well below 4, the generated vortex should remain axisymmetric  \cite{Yang2012}.

Thus, the flow field is experimentally characterized by particle image velocity (PIV) measurements with a vertical laser sheet illuminating a plane passing through the axis of the disk. To avoid parasitic light reflections, the disk is coated with black painting. The images are recorded with a high-speed camera, and the velocity fields are computed using the software DAVIS (LaVision).  For a given set of control parameter values, each individual realization shows only minor differences in the velocity fields $(u_r, u_z)$  measured in a given vertical plane ($\theta = 0$), confirming that the flow remains mainly axisymmetric, and convergence is achieved by averaging 20 independent realizations. The spatial resolution of the velocity measurement is of the order of $0.5$ mm. More details on the experimental setup can be found in \citet{Steiner_2023}.

\subsection{Numerical methods}
The flow field generated by the motion of the disk toward or away from the wall is also computed by direct numerical simulations of the Navier-Stokes equations for a Newtonian incompressible fluid using the Basilisk flow solver. The configuration is 2D axisymmetric: In the meridional plane, the computational domain is a square of width $\lambda$ defined by $(r,z)\in [0,\lambda] \times [-b-w/2,\,\lambda-b-w/2]$, where $\lambda=2D$. Changing slightly the domain size $\lambda$ does not change quantitatively the results. The solid disk is modeled by an immersed boundary. At $t=0$, the disk, which is located in the region $(r,z)\in [0,\,D/2] \times [L-w/2,\,L+w/2]$ for a disk moving toward the wall and $[0,D/2] \times [-w/2,w/2]$ for a disk moving away from the wall, is represented by a solid volume fraction. This region is then displaced in time with the velocity provided by Eq. (\ref{eq:DiskVelocity}). An axisymmetric boundary condition is set at $r = 0$, while no-slip conditions are used at the other boundaries of the domain. The spatial discretization is based on a  Cartesian grid with adaptive mesh refinement through a quadtree approach \cite{Popinet2009}. The final spatial resolution is of the order of $0.1$ mm. More details on the numerical scheme can be found in \citet{Steiner_2023}. We ensured that a higher refinement level does not change the circulation of the vortex ring more than 1\% and that the maximum time step $\delta t_\text{max}$ = $2\times10^{-5}\tau$ is small enough.

In the simulations, the ranges of stroke length $L$, disk diameter $D$, distance $b$ and kinematic viscosity $\nu$ have been extended compared to the experiments, so that $L \in [2, 16]~$cm, $D\in[5, 40]~$cm, $b\in[0.2, 13]~$cm and $\nu\in[6\times10^{-7},~9\times10^{-6}$] m$^2$/s. The range of the two dimensionless ratios $L/D \in [0.05, ~ 2]$ and $b/D\in [5\times10^{-3}, 1.3]$ has thus been extended compared to the experimental range, but the range of the Reynolds number $Re \in [1.3\times10^3, ~2\times10^4]$ has been kept the same as indicated in Table \ref{tab:Symbols}.\\

\begin{table}[b]
\centering
\setlength{\tabcolsep}{6pt}
\begin{tabular}{ |c c c c c c c c c| } 
 \hline
 $b$ (cm) & $L$ (cm) & $D$ (cm) & $\tau$ (s) & $\nu$ (m$^2$/s) & $L/D$ & $b/D$ & $Re$ & Symbols\\
 \hline
0.2 & 2 - 5.2 & 5 - 15 & 0.5 - 0.52 & $10^{-6}$ & 0.19 - 0.56 & 0.013 - 0.04 & $1.3\times10^3$ - $8.8\times10^3$ & \markerha / \markerhaSimu \\
0.5 & 2 - 5.2 & 10 & 0.5 - 2.5 & $10^{-6}$  & 0.2 - 0.52 & 0.05 & $1.8\times10^3$ - $8.8\times10^3$  & \markerhb / \markerhbSimu\\
1 & 2 - 5.2 & 5 - 15 & 0.5 - 2.5 & $10^{-6}$  & 0.19 - 0.56 & 0.067 - 0.2 & $1.8\times10^3$ - $9.8\times10^3$ & \markerhc / \markerhcSimu\\
1.5 & 2 - 5.2 & 10 & 0.5 - 2.5 & $10^{-6}$  & 0.2 - 0.52 & 0.15 & $1.8\times10^3$ - $1.3\times10^4$ & \markerhd / \markerhdSimu \\
2 & 2 - 5.2 & 7.5 - 15 & 0.36 - 0.83 & $10^{-6}$  & 0.19 - 0.52 & 0.13 - 0.27 & $5.3\times10^3$ - $2\times10^4$ & \markerhe / \markerheSimu \\
 0.2 - 13 & 2.8 - 11.2 & 10 - 40 & 0.5 & $6\times10^{-7}$ - $9\times10^{-6}$  & 0.28 & 0.005 - 1.3 & $1.5\times10^4$ & \markerhD \\
0.4 - 2 & 2 - 16 & 8-40 & 0.5 & $6\times10^{-7}$ - $2.7\times10^{-6}$ & 0.05 - 2 & 0.05 & $1.5\times10^4$& \markerLD \\
 \hline
\end{tabular}
 \caption{Sets of experimental and numerical parameters, associated non-dimensional numbers, and corresponding symbols used for figures. Empty symbols correspond to experiments, and full symbols correspond to numerical simulations.}
 \label{tab:Symbols}
\end{table}

\subsection{Vortex characterization}

A vortex ring is formed in the wake of the translating disk, and the resulting velocity fields obtained from the PIV measurements or numerical simulations are analyzed with custom MATLAB routines described in \citet{Steiner_2023}. The circulation of the vortex is computed by $\Gamma = \iint_S \omega_\theta ~\mathrm{d}r \, \mathrm{d}z$, where $\omega_\theta = \partial u_r/\partial z - \partial u_z/\partial r $ is the local vorticity. The horizontal and vertical positions of the vortex are given by $r_\mathrm{G} = \iint_S (r\, \omega/\Gamma) ~\mathrm{d} r \, \mathrm{d}z$ and $ z_\mathrm{G} = \iint_S (z\, \omega/\Gamma) ~\mathrm{d}r\, \mathrm{d}z$ and the core radius $a$ of the vortex is obtained from the second-moment method described in \citet{le_dizes_viscous_2002} and used in \citet{Steiner_2023}. An iterative routine gives the final circulation of the vortex, computed as the total circulation on a surface $S$ five times the size of the vortex. This size was chosen to include most of the circulation in the computation domain without capturing too much noise. 

In the unbounded case, where the disk is far from any boundaries ($b/a \gg 1)$, the maximum circulation $|\Gamma_m|$ and the maximum radius $a_m$ of the vortex ring formed in the wake of a translating disk whose velocity evolves sinusoidally from start to stop have been shown by \cite{Steiner_2023} to follow the scaling laws 
\begin{equation}
|\Gamma_m| \simeq 2.1 \, L^{4/3}D^{2/3} \tau ^{-1}  \quad \mathrm{and} \quad  a_m \simeq 0.1 \, L^{2/3}D^{1/3}.
\label{eq:RappelInf}
\end{equation}

These scalings laws have been shown to be in agreement with the theoretical predictions of  \citet{wedemeyer_ausbildung_1961} for the bidimensional  configuration of a plate moving at a constant velocity, but the numerical prefactors are specific to the geometry of the object and to the law of motion. The values 2.1 and 0.1 given in Eq. (3) correspond to the present specific conditions of a disk in sinusoidal translation. These scalings have been found by \citet{Steiner_2023} both experimentally and numerically with a good agreement meaning that the rod holding the disk experimentally does not have a significant influence on the results and that the flow is essentially axisymmetric.  The Reynolds number characteristic of the vortex flow can be defined as $Re_\Gamma = |\Gamma_m|/\nu$ and is linked with the disk Reynolds number $Re = V_m D/\nu$ defined above in section II.A  by the relation $Re_\Gamma =(4.2/\pi) (L/D)^{1/3} Re$. As the ratio $L/D$  does not vary so much in the present study and is of order one, the values of these two Reynolds numbers are thus very close.

In the presence of a bottom wall, the minimum distance $b$ between the disk and the wall is expected to play a significant role in the generation process. By dimensional analysis, the dimensionless coefficients for the maximum circulation $c_\Gamma$ and the maximum radius $c_a$ defined as
\begin{equation}
c_\Gamma = \frac{|\Gamma_m|}{L^{4/3}D^{2/3} \tau ^{-1}}, \quad \mathrm{and} \quad c_a = \frac{a_m}{L^{2/3}D^{1/3}},
\label{eq:AdimCoefficients}
\end{equation}
are expected to be some functions of the two ratios $b/D$ and $L/D$ instead of the constant values $c_\Gamma \simeq 2.1$ and $c_a \simeq 0.1$ corresponding to the unbounded case found in \citet{Steiner_2023}.

We first consider the case of a disk moving toward a bottom wall  in Sec. III and then the case of a disk moving away from a bottom wall in Sec. IV. In both cases, we will first describe the flow generated in the near wake of the disk and then highlight the dependence of the starting vortex on the parameters $L/D$ and $b/D$ compared to the unbounded case (see Eq. (\ref{eq:RappelInf})).

\begin{figure}[t!]
    \centering
    \includegraphics[width=\linewidth]{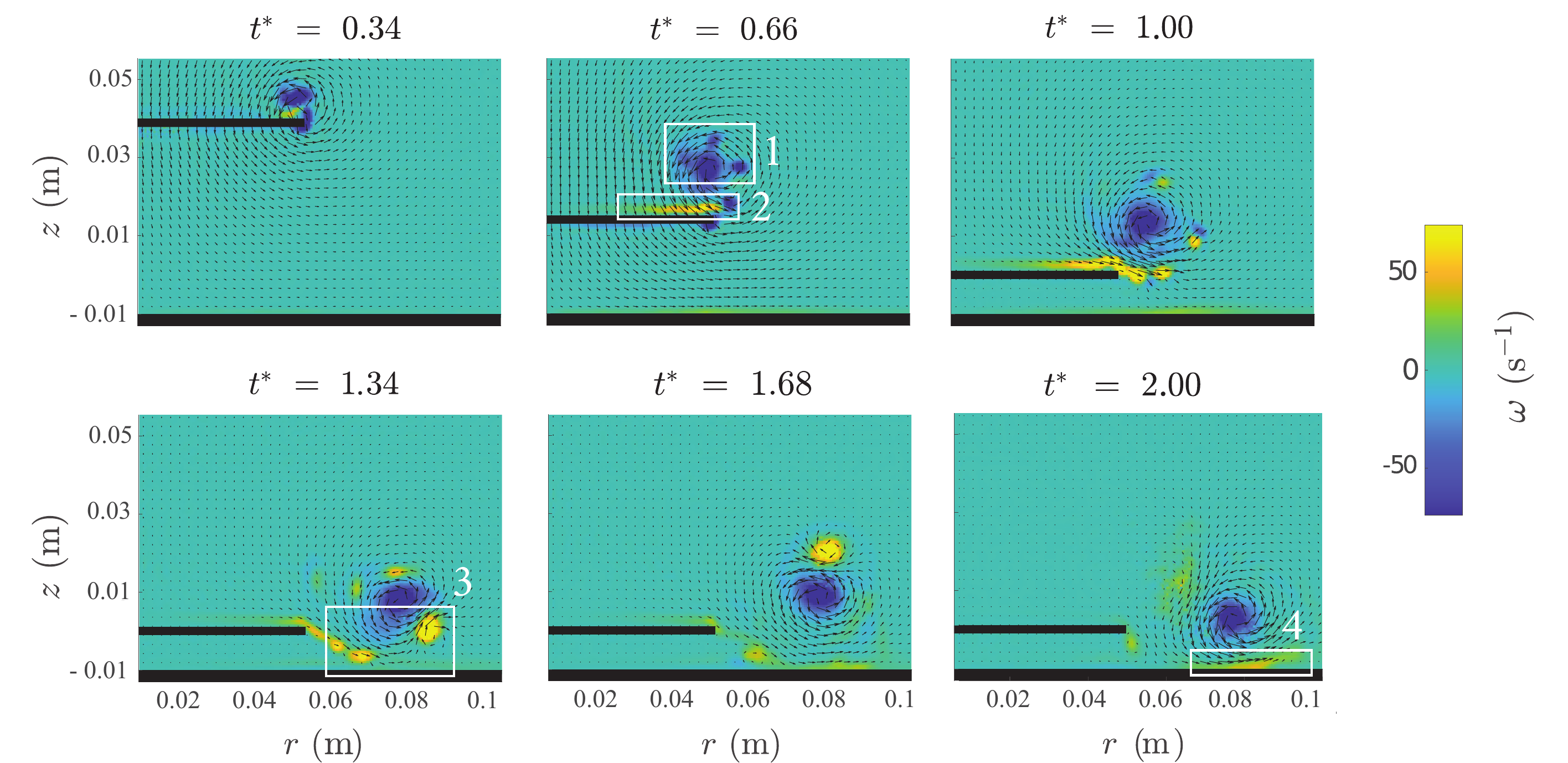}
    \caption{Velocity field (arrows) and vorticity field (color scale) generated by a disk starting from the top position and translating toward a horizontal wall in water, obtained experimentally using PIV measurements at different dimensionless times $t^*$ for $L=5.2~$cm, $D$ = 10 cm, $\tau=0.83~$s, and $b=1~$cm ($Re = 9.8 \times 10^3$, $L/D = 0.52$, $b/D = 0.1$). The lateral wall is located at $r=0.2~$m $= 4D/2$.}
    \label{fig:Fig2_Snapshot}
\end{figure}

\section{Disk translation toward the wall}
\label{SecIII}

\subsection{General flow description}

In this section, the disk moves toward the wall from the distance $b + L$ at the initial dimensionless time $t^* = 0$ to the minimum distance $b$ at $t^* = 1$. An example of the velocity (arrows) and vorticity (color scale) fields obtained experimentally for the dimensional parameters $L=5.2~$cm, $D=10~$cm, $\tau=0.83$ s, and $b=1~$cm, corresponding to the nondimensional parameters $Re = 9.8 \times 10^3$, $L/D = 0.52$, and $b/D = 0.1$, are shown in Fig. \ref{fig:Fig2_Snapshot} at different times $t^*$.

As in the infinite medium configuration where no bottom wall is present \cite{Steiner_2023}, an initial vortex ring forms in the wake of the disk. It grows and accumulates vorticity due to the detachment of the boundary layer and the winding of the vortex sheet as the disk moves; this growth is particularly visible between $t^*~=~0.34$ and $t^*~=~0.66$. It can be observed that the vortex sheet, due to Kelvin-Helmholtz instability, concentrates into several small satellite vortices (see Box 1) that rotate around the starting vortex \cite{Luchini2002,Luchini2017, higuchi_numerical_1996,Pierce1961,higuchi_numerical_1996}. At the same time, a vorticity layer of opposite sign forms between the starting vortex and the top of the disk (see Box 2 at $t^*~=~0.66$). When the disk stops at $t^*=1$, due to the circumferential velocity induced by the starting vortex, this vorticity layer is reaped off the disk, and several small stopping vortices are formed one after the other (see Box 3 at $t^*=1.34$), as in the unbounded configuration \cite{Steiner_2023}. Then, between $t^*=1.34$ and $t^*=2$, the main stopping vortex rotates around the starting vortex. In addition, the interaction of the starting vortex with the bottom of the tank leads to the formation of a new vorticity layer of opposite sign, which is barely visible at $t^*=1.68$ but clearly visible at $t^*=2$ (see Box 4). The formation of such a vorticity layer has already been observed when a vortex ring hits a wall \cite{walker_impact_1987, orlandi_vortex_1990, orlandi_vortex_1993, swearingen_dynamics_1995}. Finally, the motion of the starting vortex is largely constrained by the presence of the wall after the disk stops, causing the vortex to bounce off the wall.

In addition to the generation of the vortex rings, some fluid is also radially ejected from the gap between the wall and the disk as the disk moves. As can be observed in Fig. \ref{fig:Fig2_Snapshot} at $t^*=0.66$, the flow is essentially parallel to the wall. After the disk stops ($t^*>1$), the radial flow stops. As we have just described, the flow is then essentially composed of vortex rings interacting with each other and with the wall. 

\subsection{Outward radial flow in the disk-wall gap}
We first focus on the features of the outward radial flow stemming from the axial squeezing of the fluid region between the disk and the wall.
\begin{figure}[t!]
     \centering
     \includegraphics[width=0.7\linewidth]{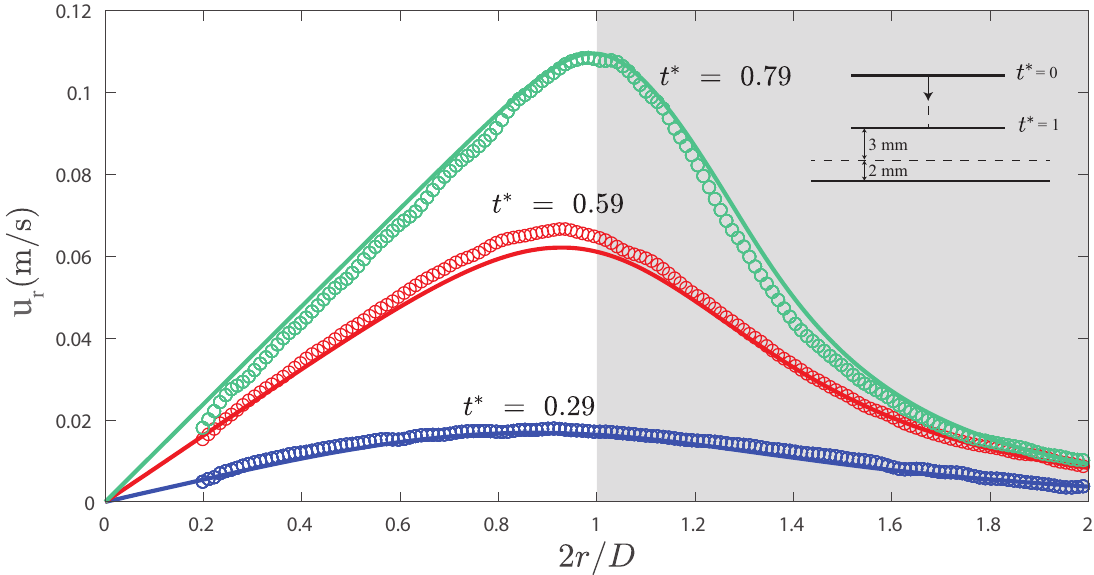}
     \caption{Radial distribution of the radial velocity $u_r$ at $2~$mm from the bottom wall ($z = -3~$mm) at three successive times in the case of a disk of diameter $D = 10$ cm moving toward the wall for $L = 4.4$ cm, $\tau = 1$ s and $b = 0.5$ cm ($Re = 6.9 \times 10^3$, $L/D = 0.44$, $b/D = 0.05$). The open symbols (\textcolor{blue}{o}) correspond to experiments, and the solid lines correspond to numerical simulations. The grey zone corresponds to radial positions outside of the disk-wall gap $(2r/D > 1)$.} 
\label{fig:Fig2bis_HorizontalVelocityDPH_RadialDependency}
 \end{figure}
An example of the variation of the radial velocity as a function of $r$ at $z = -3~$mm, i.e. $2~$mm from the wall, is shown in figure \ref{fig:Fig2bis_HorizontalVelocityDPH_RadialDependency} for different times and for $L$ = 4.4 cm, $D$ = 10 cm, $\tau$ = 1 s and $b = 0.5~$cm ($Re = 6.9 \times 10^3$, $L/D = 0.44$, $b/D = 0.05$). Good agreement can be observed at all times between the numerical simulations and the experimental measurements. The radial velocity of the flow within the gap between the disk and the wall increases with $r$ from zero at the center of the disk up to a maximum near the edge of the disk at $r = D/2$ and then decreases outside the gap. In the following, we focus on the maximum radial velocity at $r = D/2$.

 \begin{figure}[h!]
     \centering
     \includegraphics[width=\linewidth]{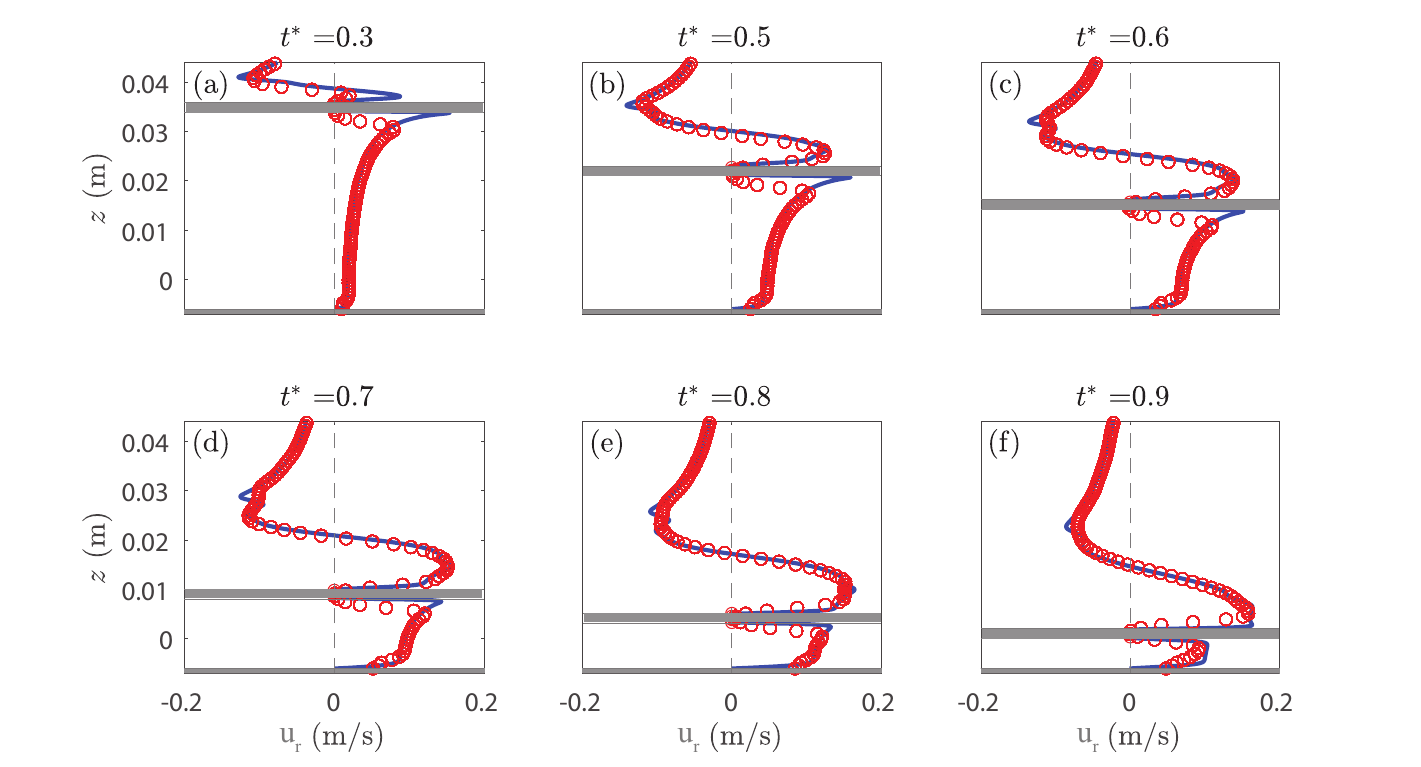}
     \caption{Vertical distribution of the radial velocity  $u_r(z)$ at $r = D/2$ and different time $t$ for the same flow configuration as in Fig. 3. The upper (resp. lower) horizontal thick gray line corresponds to the disk (resp. bottom wall) region. The red symbols correspond to the experimental measurements, and the blue solid lines correspond to the numerical simulations.}
     \label{fig:FigXX_HorizontalVelocityDPH}
\end{figure}

An example of the vertical distribution of the radial velocity $u_r(z)$ at $r=D/2$ is shown in figures \ref{fig:FigXX_HorizontalVelocityDPH}(a)-(f) at different times. There is again a good agreement between the simulation and the experiment away from any solid walls, i.e. away from the disk and the bottom of the tank. However, in the immediate vicinity of these walls, experiment and simulation are no longer in perfect agreement. Indeed, the experimental PIV measurements are not resolved well enough to capture the flow in the boundary layers since the thickness of the boundary layer is of the order of a millimeter, while the resolution of the experiments is of the order of $0.5$ mm compared to $0.1~$mm in the simulations. Experimentally, therefore, there are only very few velocity vectors in the boundary layer. However, the good agreement between the simulation and the experiment away from the walls suggests that the simulations are reliable and precisely account for the flow at the walls. The three boundary layers at the bottom of the tank and at both sides of the disk are most visible in the simulation.

The radial velocity profiles $u_r(z)$ above the disk in Figs. 4(a) to (f) are typical of a Lamb-Oseen type vortex. The radial velocity profiles $u_r(z)$ under the disk in Figs. 4(a) to (f) show that the fluid is expelled from the gap between the disk and the wall as expected, with a maximum velocity near the disk from which the fluid escapes before rolling up into the vortex. At the last instant shown here ($t^* = 0.9$), the velocity is still far from a parabolic profile, which would correspond to the lubrication assumptions for a very thin gap. Indeed, the viscous time $h^2(\tau)/\nu$ is 25 s whereas the inertial time $D/U$ is around 1 s for the case of Fig. 4, where $U$ is the mean horizontal velocity between the wall and the disk. As $h^2(\tau)/\nu > D/U$, the viscous forces are much smaller than the inertial forces, so that the lubrication approximations cannot be used here to solve the flow in the gap between the disk and the wall.

 \begin{figure}[h!]
     \centering
     \includegraphics[width=\linewidth]{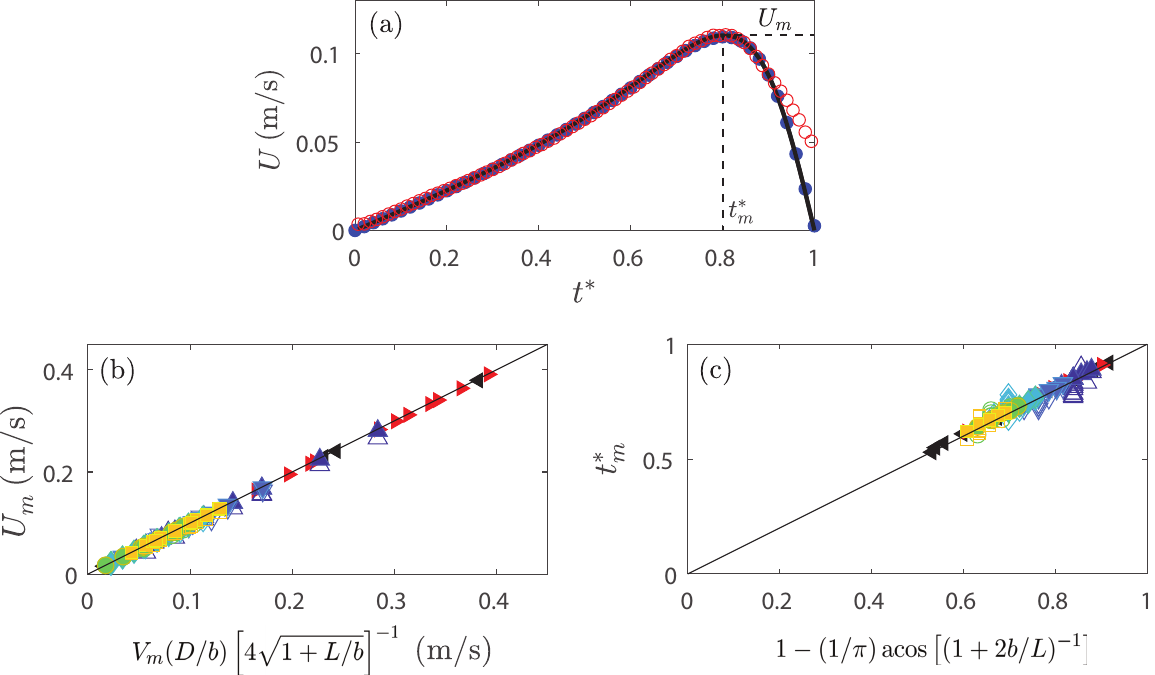}
     \caption{(a) Time evolution of the mean radial $U$ velocity at $r = D/2$ between the disk and the wall for the case of Fig. 3. Open symbols correspond to the experiments, filled symbols to the numerical simulations, and the solid line to the analytical expression given by Eq. (\ref{eq:radial_velocity_toward}). (b) Maximum radial velocity at $r = D/2$ measured from experiments (open symbols) and simulations (filled symbols) versus the analytical prediction given by Eq. (\ref{eq:maximum_radial_velocity_toward}). (c) Time for the maximum radial velocity measured in the experiments and numerical simulations versus the analytical prediction given by Eq. (\ref{eq:maximum_time_toward}).}
     \label{fig:FigXX_Towards_UMeanAndTimeMaxDPH}
\end{figure}

To model the present inertial outward flow, we consider the continuity equation $(1/r)[\partial \big(r u_r(r,z,t)\big)/\partial r] + \partial u_z(r,z,t)/\partial z = 0$ which leads by integration over the gap to

\begin{equation}
    \frac{1}{r}\frac{\partial \big(r U(r,t)\big)}{\partial r} + \frac{1}{h(t)}\frac{dh(t)}{dt} = 0,
\end{equation}

where $U(r,t) = \int^{h(t)}_0 u_r(r,z,t)~\mathrm{d}z/h(t)$ is the gap-averaged radial velocity. With a last integration over $r$ considering the boundary condition $U= 0$ at $r=0$, this ultimately leads to the relation

 \begin{equation}
     U(r,t) = -\frac{r}{ 2h(t)} \frac{dh(t)}{dt}.
     \label{eq:general_radial_velocity}
 \end{equation}

The gap-averaged radial velocity $U$ is thus related to the disk velocity $dh/dt$ by the gap thickness $h$ and the radial position $r$. Considering the motion law for the disk given by Eqs (\ref{eq:DiskPosition})-(\ref{eq:DiskVelocity}), the gap-averaged radial flow $U$ is thus
 \begin{equation}
     U(r,t) = \frac{r V_m \sin (\pi t^*)}{2b + L[1+\cos (\pi t^*)]},
     \label{eq:radial_velocity_toward}
 \end{equation}
for $0 \leqslant t^*\leqslant 1$. According to Eq. (\ref{eq:radial_velocity_toward}), this gap-averaged flow should be maximal at the edge of the disk ($r=D/2$) and should reach a maximum value $U_m$ at the time $t^*_m$, which are respectively given by  

\begin{equation}
     U_m = \frac{D}{b}\frac{V_m}{4 \sqrt{1+L/b}},
     \label{eq:maximum_radial_velocity_toward}
\end{equation}
\begin{equation}
     t^*_m = 1 - \frac{1}{\pi} \, \text{acos}\left(\frac{1}{1+2b/L}\right).
     \label{eq:maximum_time_toward}
\end{equation}

An example of the time evolution of the mean radial velocity $U(D/2,~t)$ obtained experimentally and numerically is shown in Figure \ref{fig:FigXX_Towards_UMeanAndTimeMaxDPH}(a) and compared with the prediction from Eq. (\ref{eq:radial_velocity_toward}). The average velocity $U(D/2,~t)$ first increases before decreasing after reaching a maximum value of $U_m \simeq 0.11~$ m/s during the second part of the motion at time $t^*_m \simeq 0.8$. In the following, $U_m$ will be referred to as the maximum ejection velocity. The numerical and theoretical velocities $U(D/2,~t)$ are very similar, and there is also a very good agreement with the experiment up to $t^* \simeq 0.9$. The discrepancies observed for $t^* \gtrsim 0.9$ can be explained by the fact that it becomes increasingly difficult to obtain an accurate experimental measurement of the mean flow between the disk and the wall by PIV means. Indeed, at such a late time, the disk approaches the wall very close so that fewer and fewer experimental velocity vectors are computed. For example, in the case presented here, the final distance between the disk and the wall is $b = 0.5~$cm while the experimental resolution for the velocity vectors is 0.06 cm, so the average is calculated from only 8 values. 

Figure 5(b) shows the maximum ejection velocity $U_m$ measured in each experiment (empty symbols) and simulation (solid symbols) plotted against the predicted value $U_m$ from Eq. (\ref{eq:maximum_radial_velocity_toward}). The velocities predicted by the model show a very good agreement with the simulations and the experiments for the entire range of parameters considered here. Finally, the time $t^{*}_m$ at which the maximum ejection velocity is reached for all the experiments and simulations is plotted in Fig. 5(c) against the predicted time $t^*_m$ from Eq. (\ref{eq:maximum_time_toward}). Again, we see a very good agreement between the experiments and the numerical simulations with the predicted value. All time values are between 1/2 and 1, as expected from Eq. (\ref{eq:maximum_time_toward}), since $1+2b/L >1$.

\subsection{Features of the starting vortex}
 We now consider the key features of the starting vortex in the wake of the disk using both experiments and simulations. We will first present the time dynamics of the vortex formation and then the scaling laws of its circulation and size with the control parameters. 
\subsubsection{Time Dynamics}

\begin{figure}[t!]
    \centering
    \includegraphics[width=\linewidth]{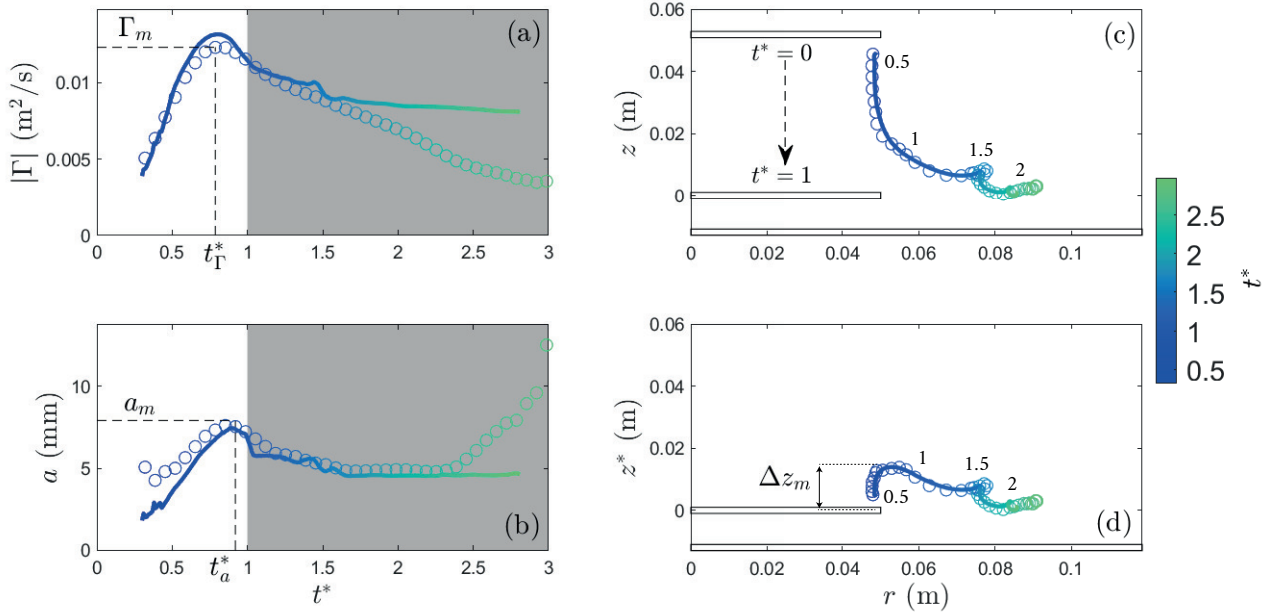}
    \caption{Time evolution of (a) the circulation $|\Gamma|$, (b) the radius $a$, and (c),(d) the position ($r_G$, $z_G$) of the starting vortex in the frame of reference of (c) the laboratory and (d) the disk for the same case as in Fig. 2. The symbols (\protect\markerExpEvolution) represent experimental data, and the solid lines (\textcolor{blue}{\textbf{-}}) represent the corresponding numerical simulation data. The color bar indicates the dimensionless time $t^*$.}
    \label{fig:FigXX_VortexFeaturesDPHFlipped_To3H}
\end{figure}
Figure \ref{fig:FigXX_VortexFeaturesDPHFlipped_To3H} reports an example of the temporal evolution of the circulation [Fig. \ref{fig:FigXX_VortexFeaturesDPHFlipped_To3H}(a)], the radius [Fig. \ref{fig:FigXX_VortexFeaturesDPHFlipped_To3H}(b)] and the position of the vortex in the reference frame of the laboratory [Fig. \ref{fig:FigXX_VortexFeaturesDPHFlipped_To3H}(c)] or the disk [Fig. \ref{fig:FigXX_VortexFeaturesDPHFlipped_To3H}(d)].
As previously observed in an infinite medium \cite{Steiner_2023}, three phases of vortex evolution are observed. The first phase corresponds to the formation of the vortex, during which the circulation increases and the vortex grows in size ($t^*\lesssim 0.8$). At the same time, the vortex starts to move away from the disk, as can be seen in Fig. \ref{fig:FigXX_VortexFeaturesDPHFlipped_To3H}(d). The vortex reaches its maximum circulation $|\Gamma_m| = 1.2\times10^{-2}~$m$^2$/s at $t^*_{\Gamma} \simeq 0.78$ and its radius grows up to $a_m = 7.6~$mm at $t^*_{a} \simeq 0.85$ in the example reported here. Both times $t^*_{\Gamma}$ and $t^*_{a}$ are close to the time $t^*_m \simeq 0.8$ at which the outward radial flow from the gap is maximal. In an infinite medium, Eq. (\ref{eq:RappelInf}) would predict a maximum circulation of $|\Gamma_m| \simeq 10^{-2}$ m$^2$/s and a maximal radius of $a_m \simeq 6.2~$mm slightly smaller than what is observed here, meaning that the confinement increases the energy injected into the vortex ring.

During the second phase ($0.8 \lesssim t^* \leqslant 1 $), the circulation and radius decrease with time due to the interaction between the vortex and the decelerating disk. As in an infinite medium, this decrease of the circulation is attributed to the entry of oppositely signed vorticity located above the disk into the vortex (see Box 2 in Fig. 2(a) at $t^*=0.66$). In addition, during this second phase, the starting vortex moves radially outwards due to the deceleration of the disk. The outward radial motion of the vortex core corresponds to an increase of  the radius of the vortex ring which leads to a decrease of the radius of the vortex core by volume conservation.

\begin{figure}[t!]
    \centering
    \includegraphics[width=\linewidth]{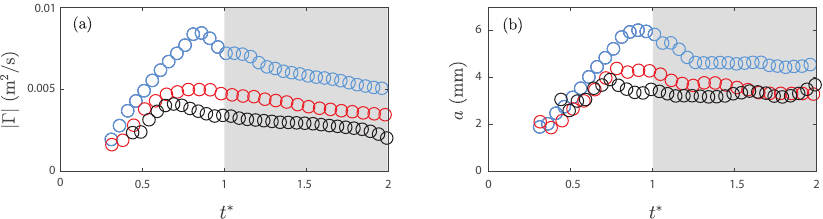}
    \caption{Time evolution of (a) the circulation $|\Gamma|$ and (b) the core radius $a$ of the starting vortex measured experimentally for a disk moving toward a wall for $L=2.8~$cm, $D=10~$cm and $\tau=0.83~$s ($Re = 5296$, $L/D = 0.28$), and for the stopping distance $b=0.2$ cm $=0.02D$ (blue), $b = 1$ cm $=0.1D$ (red), $b = 20$ cm $=2D$ (black).}
    \label{fig:Fig4_Towards_VortexFeatures_DifferentHeights}
\end{figure}

Finally, the third phase of the vortex evolution begins after the disk stops ($t^*>1$). The vortex is no longer fed by the disk motion and detaches from the disk, while the radius of the ring also continues to increase. The starting vortex hits the bottom wall in a way similar to what has been observed in previous studies \cite{walker_impact_1987, orlandi_vortex_1990, orlandi_vortex_1993, swearingen_dynamics_1995, kramer_vorticity_2007, lim_experimental_1989}. The starting vortex exhibits a curved trajectory due to the influence of the stopping vortex, as can be seen in Figs. \ref{fig:FigXX_VortexFeaturesDPHFlipped_To3H}(c),(d).  The radius $a$ of the vortex slowly decreases with time for $1<t^* \lesssim 2$ [Fig. \ref{fig:FigXX_VortexFeaturesDPHFlipped_To3H}(b)]. The decrease in the radius of the core vortex after the disk stops can be explained by the increase in the radius of the ring and volume conservation as in an infinite medium \cite{Steiner_2023}. Finally, the circulation also gradually decreases for $t^* > 1$ as seen in Fig. \ref{fig:FigXX_VortexFeaturesDPHFlipped_To3H}(a).

The numerical results are in relatively good quantitative agreement with the experimental ones for $t^*\lesssim~2$. However, beyond this time, the circulation of the vortex obtained from the numerical simulation becomes significantly larger than the experimental one. Moreover, after $t^*=2$, the radius of the ring obtained experimentally grows, while the size of the ring obtained numerically remains almost constant (see the last symbols in Fig. 6(b)). This divergence is probably due to the destabilization of the starting vortex after its impact on the bottom wall, causing it to lose its axisymmetry, which, on the contrary, is forced in the simulations. This deviation is consistently observed in all the cases investigated here.

Examples of the time evolution of the circulation and the radius of the starting vortex for three different minimum distances $b$ from the bottom wall are shown in Figs. \ref{fig:Fig4_Towards_VortexFeatures_DifferentHeights}(a) and (b), respectively. The black symbols correspond to $b= 20~$cm, which can be considered as an infinite medium case ($b/a \gg 1$). We observe that the maximum circulation and the radius increase significantly as $b$ decreases. In particular, the circulation is doubled between $b\simeq20~$cm and $b=0.2~$cm, while the radius increases by 50\%. The time at which the maximum circulation and radius are reached suggests that the closer the disk approaches the wall, the longer the duration during which the circulation and radius increase. This can be understood because, in the gap between the disk and the wall, the fluid is ejected at a higher velocity as the disk gets closer to the wall, which sustains vorticity ejection into the vortex. This is encoded in Eq. (\ref{eq:maximum_time_toward}), which gives the evolution of $t^*_m$ with $b$. The changes in circulation and radius of the vortex highlight the ground effect that has been described in the introduction.

\subsubsection{Scaling laws}
\begin{figure}[t!]
    \centering
    \includegraphics[width=0.8\linewidth]{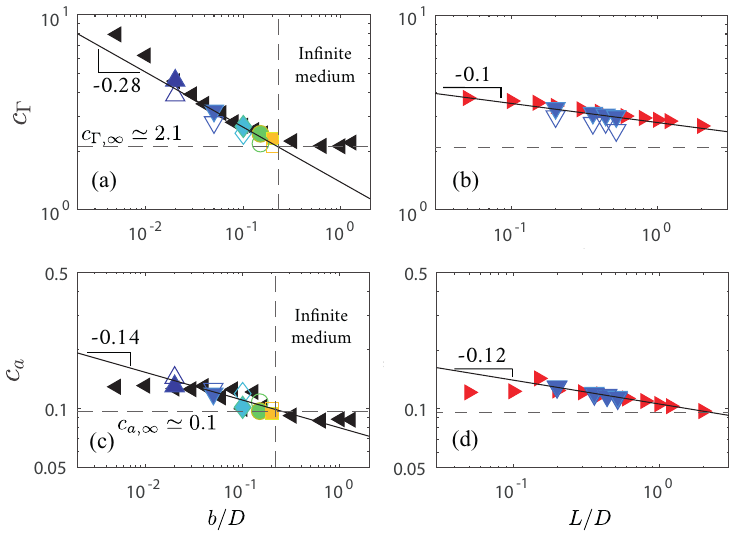}
    \caption{(a) Dimensionless maximum circulation $c_\Gamma$, (c) dimensionless maximum radius $c_a$, as a function of $b/D$ for $L/D = 0.28$, and (b) $c_\Gamma$, (d) $c_a$ as functions of $L/D$ for $b/D = 0.05$ for a disk moving toward a wall. The horizontal dashed lines represent the values in an infinite medium: (a),(c) $c_{\Gamma, \infty} \simeq 2.1$, and (b),(d) $c_{a, \infty} \simeq 0.1$. The solid lines are given by the equations (a) $c_\Gamma = 1.4 (b/D)^{-0.28}$, (b) $c_\Gamma = 2.8 (L/D)^{-0.1}$, (c) $c_a = 0.08 (b/D)^{-0.14}$ and (d) $c_a = 0.1 (L/D)^{-0.12}$.}
    \label{fig:Fig5_Towards_cAll_h0D_LD}
\end{figure}

For large values of $b$, \textit{i.e.}, in the unbounded case, and in the inertial regime where the effect of $Re$ can be neglected, the two dimensionless coefficients $c_\Gamma$ and $c_a$ defined in Eq. (\ref{eq:AdimCoefficients}) for the maximum circulation and core radius of the vortex are constant \cite{Steiner_2023}. As the disk moves toward a wall, one can expect these coefficients to depend on both dimensionless parameters $b/D$ and $L/D$.
To investigate the effect of $b/D$ and $L/D$ independently, we have varied $b/D$ keeping $L/D$ constant and varied $L/D$ with constant $b/D$ (sixth and seventh rows of Table \ref{tab:Symbols}).

The dimensionless coefficients $c_\Gamma$ and $c_a$ defined in Eq. (\ref{eq:AdimCoefficients}) are plotted as a function of $b/D$ and $L/D$ in Fig. \ref{fig:Fig5_Towards_cAll_h0D_LD}. The black and red symbols correspond to a Reynolds number of $Re=1.5\times10^4$, while the other colored dots correspond to Reynolds numbers ranging from $3\times10^3$ to $9\times10^3$. The Reynolds number does not affect the results in the range of parameters considered here, confirming that viscous effects do not play a significant role in the vortex properties in the present inertial regime.

In Fig. \ref{fig:Fig5_Towards_cAll_h0D_LD}(a) for $b/D ~\gtrsim 0.2$, the dimensionless circulation $c_\Gamma$ is found to be independent of $b/D$ and equal to the unbounded value $c_{\Gamma,\infty} \simeq 2.1$. However, we observe that $c_\Gamma$ increases with decreasing $b/D$ for $b/D \lesssim 0.2$, illustrating the ground effect. A power law $c_\Gamma \sim (b/D)^{-0.28}$ fits the data well for $b/D ~\lesssim 0.2$ over a decade. However, two points where $b/D \lesssim 10^{-2}$, i.e., of very strong confinement, deviate from the power law. For these points $b^2/\nu < 5 \tau$, suggesting that viscous effects between the disk and the wall are likely to become significant. In Fig. \ref{fig:Fig5_Towards_cAll_h0D_LD}(b), we observe that $c_\Gamma$ decreases with increasing $L/D$ according to the power law $c_\Gamma~\sim~(L/D)^{-0.1}$. Thus, for fixed values of $D$ and $b$, the influence of the wall is less pronounced for larger stroke lengths $L$. This observation comes from the fact that, if the ratio $L/b$ is very large, the disk, and thus the vortex, will evolve over a longer distance without the influence of the wall. On the opposite, if $L/b$ is small, the effect of the wall is felt throughout the disk's stroke and plays a significant role during all the generation of the vortex.

The dimensionless vortex radius $c_a$ is plotted as a function of $b/D$ and $L/D$ in Figs. \ref{fig:Fig5_Towards_cAll_h0D_LD}(c) and (d), respectively. As for $c_\Gamma$, we observe in Fig. \ref{fig:Fig5_Towards_cAll_h0D_LD}(c) that $c_a$ does not depend on $b/D$ for $b/D~\gtrsim0.2$ and is close to the unbounded value $c_a\simeq0.1$, while it increases with decreasing $b/D$ for $b/D \lesssim 0.2$. A power law $c_a\sim(b/D)^{-0.14}$ fits the data quite well for $b/D~\lesssim 0.2$ except for the two points of smallest $b/D$ ($b/D\lesssim10^{-2}$) where, as discussed above, viscous effects may appear. In Fig. \ref{fig:Fig5_Towards_cAll_h0D_LD}(d), we observe that $c_a$ decreases with increasing $L/D$ according to the power law $c_a\sim(L/D)^{-0.12}$, as the influence of the wall on the radius of the vortex is less pronounced for larger stroke length $L$.

In summary, the values of the non-dimensional coefficients $c_\Gamma$ and $c_a$ obtained for a disk moving toward a wall are both larger than those obtained in an infinite medium. In such a case, the scaling of these two coefficients is $c_\Gamma \propto (L/D)^{-0.1}(b/D)^{-0.28}$ and $c_a \propto (L/D)^{-0.12}(b/D)^{-0.14}$ for the set of simulations and experiments corresponding to varying $b/D$ at $L/D$ = 0.28 and varying $L/D$ at $b/D$ = 0.05.  

\newcommand{\markercGammaInfinityFull}{\raisebox{0.5pt}{\tikz{\node[draw,scale=0.6,circle,fill=color_violet,draw=color_violet](){};}}}
  \newcommand{\markercGammaInfinityEmpty}{\raisebox{0.5pt}{\tikz{\node[draw,scale=0.6,circle,draw=color_violet](){};}}}\begin{figure}[t!]
    \centering
    \includegraphics[width=0.9\linewidth]{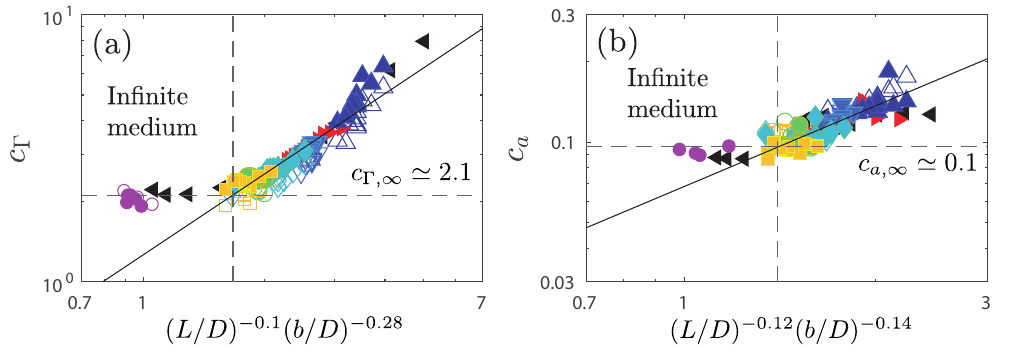}
    \caption{Maximum dimensionless (a) circulation $c_\Gamma$ as a function of $(L/D)^{-0.1}(b/D)^{-0.28}$, and (b) maximum dimensionless radius $c_a$ as a function of $(L/D)^{-0.12}(b/D)^{-0.14}$ for a disk moving toward a wall. The horizontal lines represent the values $c_{\Gamma, \infty} \simeq 2.1$ and $c_{a, \infty} \simeq 0.1$ obtained in an infinite medium \cite{Steiner_2023}. The equations of the lines are (a) $c_\Gamma = 1.3 (b/D)^{-0.28}(L/D)^{-0.1}$ and (b) $c_a = 0.07 (b/D)^{-0.14}(L/D)^{-0.12}$. All the open (resp. filled) symbols correspond to the experiments  (resp. simulations) reported in Table I except the symbols (\protect\markercGammaInfinityFull,\protect\markercGammaInfinityEmpty) that correspond to experimental and numerical results obtained by \cite{Steiner_2023} in quasi-infinite medium ($b \simeq 20~$cm).}
    \label{fig:Fig6_Towards_cMax}
\end{figure}

To test these scaling laws for all the different configurations of different $L$, $D$, and $b$ reported in Table I, $c_\Gamma$ and $c_a$ are plotted in Fig. \ref{fig:Fig6_Towards_cMax}(a) and (b) as a function of $(L/D)^{-0.1}(b/D)^{-0.28}$ and $(L/D)^{-0.12}(b/D)^{-0.14}$, respectively. All experimental and numerical data collapse well on master curves, showing a plateau corresponding to the infinite medium case followed by an increasing trend, a signature of the presence of the bottom wall. The evolution of the coefficients $c_\Gamma$ and $c_a$  with $L/D$ and $b/D$ are given by

\begin{equation}
c_\Gamma \simeq \left\{ 
\begin{array}{l}
2.1 \\
\\
1.3 \, (L/D)^{-0.1}(b/D)^{-0.28}
\end{array}
\quad \mathrm{for} \quad (L/D)^{-0.1}(b/D)^{-0.28} \, \left\{
\begin{array}{l}
\lesssim 1.6 \\
\\
\gtrsim 1.6
\end{array}
\right.
\right.
\end{equation}

\begin{equation}
c_a \simeq \left\{ 
\begin{array}{l}
0.1 \\
\\
6.8 \times10^{-2} (L/D)^{-0.12}(b/D)^{-0.14}
\end{array}
\quad \mathrm{for} \quad (L/D)^{-0.12}(b/D)^{-0.14} \, \left\{
\begin{array}{l}
\lesssim 1.4 \\
\\
\gtrsim 1.4
\end{array}
\right.
\right.
\end{equation}

The maximum circulation $\Gamma_m$ and radius $a_m$ of the starting vortex in the near-wall configuration can be rewritten in dimensional form as

\begin{equation}
  |\Gamma_m| \simeq 1.3 \, L^{1.23}D^{1.05}b^{-0.28}\tau^{-1}\quad \mathrm{and} \quad a_m \simeq 6.8\times 10^{-2}L^{0.55}D^{0.59}b^{-0.14}.
  \label{eq:circulation_radius_toward}
\end{equation}

The circulation $\Gamma_m$ and the radius $a_m$ of the vortex core are thus found to be larger when the disk moves toward a wall than when it translates in an infinite medium. They have a slightly different scaling, with a new contribution from the distance of the disk to the bottom wall $b$ with the scaling exponent $-0.28$ for $\Gamma_m$ and $-0.14$ for $a_m$, compared to the scalings recalled by Eq. (\ref{eq:RappelInf}) for the unbounded case. The effect of the stroke length $L$ on the circulation of the vortex ring seems slightly weaker with a weaker scaling exponent (1.23 instead of 4/3 in infinite medium), and the influence of $D$ is much larger with a larger scaling exponent (1.05 instead of 2/3). The effect of $L$ on the radius of the vortex is also slightly smaller with a smaller scaling exponent (0.55 instead of 2/3), and the effect of $D$ is also much larger with a larger scaling exponent (0.59 instead of 1/3). Due to the small gap between the disk and the wall, the fluid is ejected at a higher velocity than in an infinite medium. Therefore, the winding speed of the vortex ring increases. As a result, more fluid is injected into the starting vortex, which grows larger and accumulates more circulation.

\section{Disk translation from the wall}
\label{SecIV}
\subsection{General flow description}

\begin{figure}[t!]
    \centering
    \includegraphics[width=0.9\linewidth]{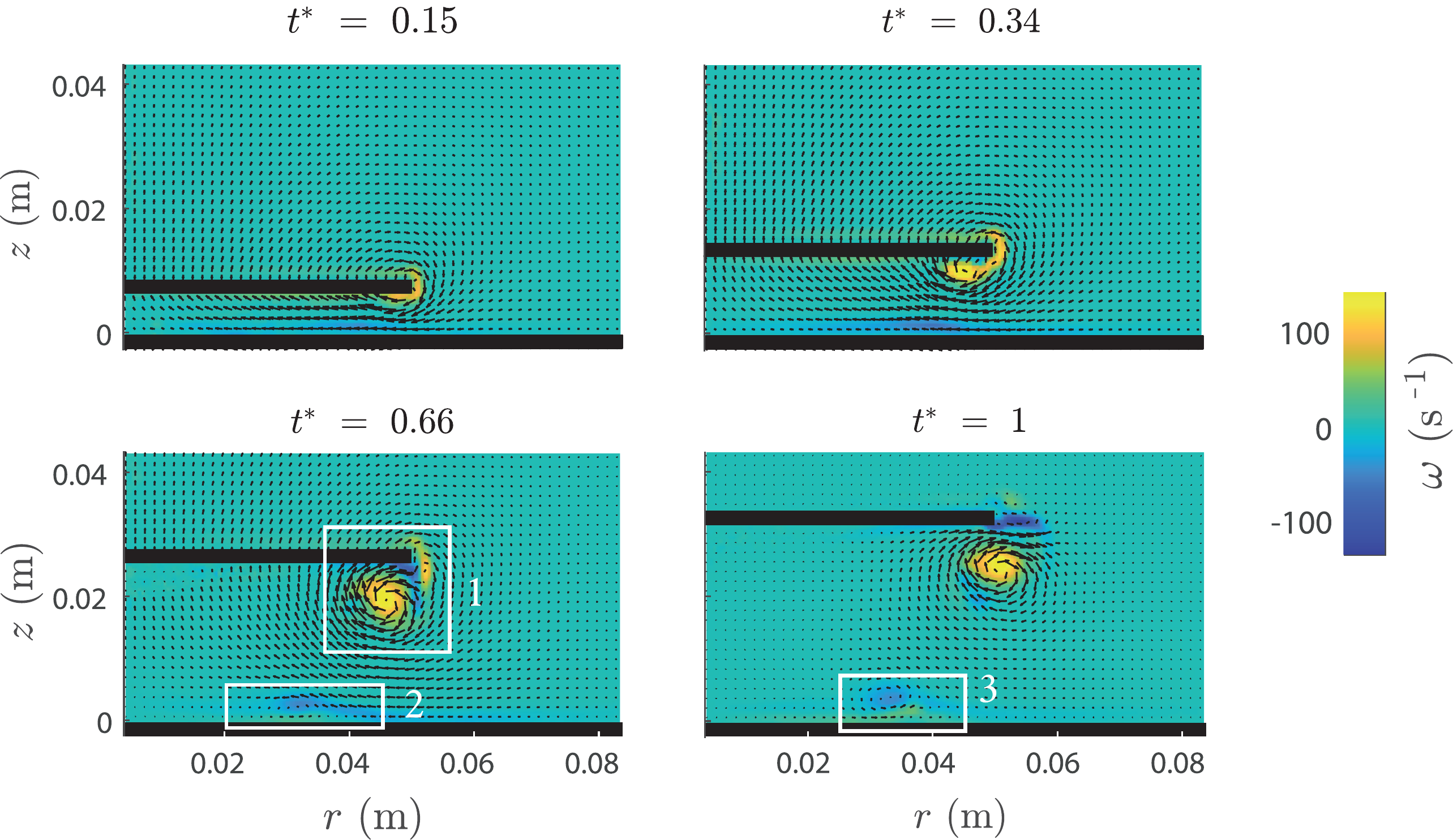}
    \caption{Velocity field (arrows) and vorticity field (color scale) generated by a disk moving away from a wall, obtained experimentally using PIV at different times $t^*$ for $D$ = 10 cm, $L=2.8~$cm, $\tau=0.83~$s, and $b=0.5~$cm ($Re = 5.3 \times 10^3$, $L/D = 0.28$, $b/D = 0.05$). The bottom wall and the disk appear in black.}
    \label{fig:FigXX_SnapshotExpAway}
\end{figure}

We now focus on the opposite configuration, where the disk is translating away from a bottom wall. An example of velocity and vorticity fields obtained experimentally by PIV for $L=2.8~$cm, $D=10~$cm, $\tau=0.83~$s, and $b=0.5~$cm ($Re = 5.3 \times 10^3$, $L/D = 0.28$, $b/D = 0.05$) is shown in Fig. \ref{fig:FigXX_SnapshotExpAway} at four different times from $t^* = 0.15$ to 1. As in the previous section, an initial vortex ring forms in the wake of the disk, which grows and accumulates vorticity due to the detachment of the boundary layer and the winding of the vortex sheet as the disk moves. At $t^*=0.66$, this vortex is clearly visible but a vorticity sheet of opposite sign has also  formed between the vortex and the disk (Box 1). In addition, a slight detachment of the boundary layer near the wall at $r\simeq3.5~$cm is observed due to the starting vortex (Box 2). This detachment leads to the formation of a new small vortex of opposite sign at the bottom of the tank (Box 3 at $t^*=1$). Note that this small vortex does not form for all parameter values, especially for larger values of $b$.  For small values of $b$, the strong inward radial flow between the disk and the wall is expected to influence the properties of the starting vortex ring. We will thus first consider the features of the inward radial flow between the disk and the wall and then those of the starting vortex ring.

\subsection{Inward radial flow in the disk-wall gap}

\begin{figure}[h!]
     \centering
     \includegraphics[width=0.7\linewidth]{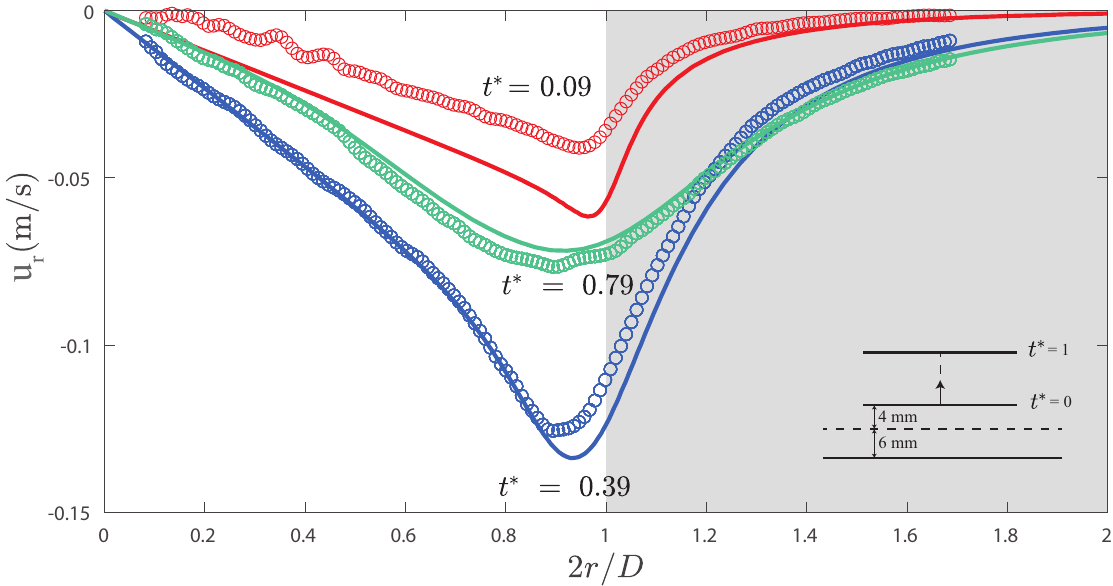}
     \caption{Radial distribution of the radial velocity $u_r$ at $6~$mm above the bottom wall ($z = -4~$mm) at three reduced times $t^* = 0.09, 0.39$ and 0.79 for a disk moving away from a bottom wall with $D = 10$ cm, $L = 2.8$ cm, $\tau = 0.5$ s and $b = 1$ cm ($Re = 8.8 \times 10^3$, $L/D = 0.28$, $b/D = 0.1$). The open symbols (o) correspond to experiments, and the solid lines (---) to numerical simulations.} 
\label{fig:FigXX_HorizontalVelocityDPB_RadialDependency}
 \end{figure}

\begin{figure}[t!]
     \centering
     \includegraphics[width=\linewidth]{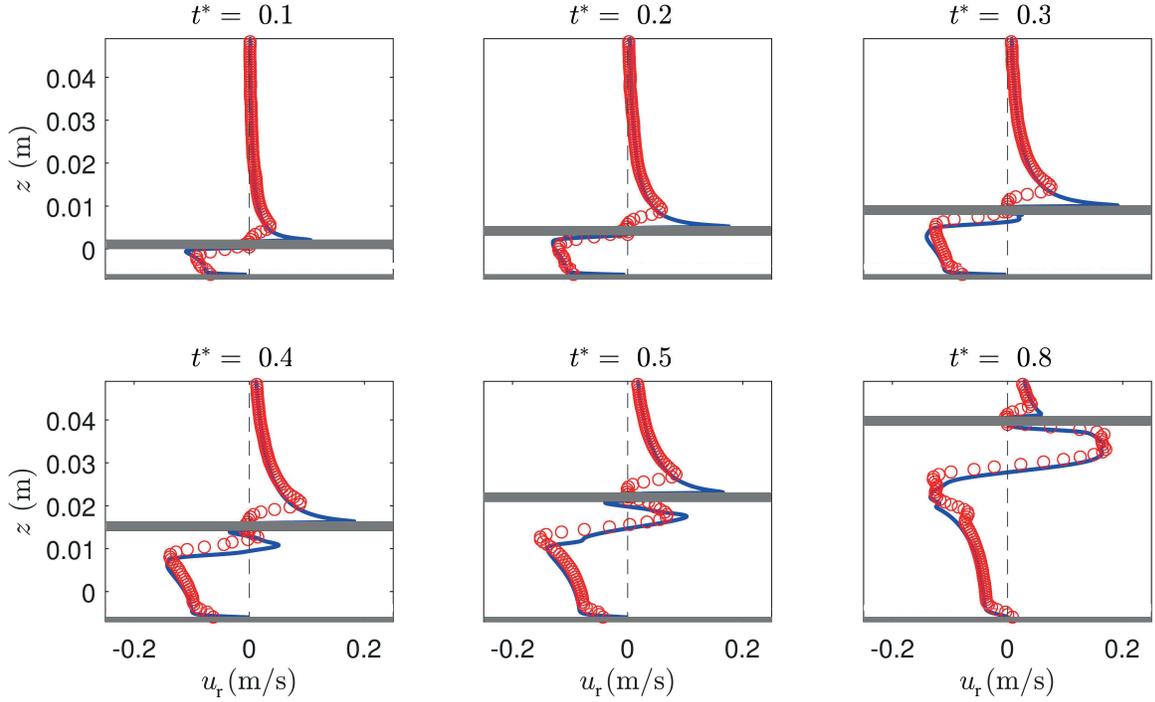}
     \caption{Radial velocity distribution $u_r(z)$ at $r = D/2$ in the case of a disk of diameter $D = 10$ cm moving away from the bottom wall for $L = 4.4$ cm, $\tau = 1$ s and $b = 0.5$ cm ($Re = 6.9 \times 10^3$, $L/D = 0.44$, $b/D = 0.05$) at different times $t^*$. The upper (resp. lower) horizontal thick gray line corresponds to the disk (resp. bottom wall). The red symbols correspond to the experimental measurements, and the blue solid lines correspond to the numerical simulations.}
     \label{fig:FigXX_HorizontalVelocityDPB}
\end{figure}

We first focus here on the characteristics of the inward radial flow stemming from the axial stretching of the fluid region between the disk and the bottom wall. An example of the variation of the inward velocity as a function of the radial coordinate $r$ at $z = -4$ mm, i.e. 6 mm from the bottom wall, for $L = 2.8$ cm, $D = 10$ cm, $\tau = 0.5$ s and $b = 1$ cm ($Re = 8.8 \times 10^3$, $L/D = 0.28$, $b/D = 0.1$) is shown in Fig. \ref{fig:FigXX_HorizontalVelocityDPB_RadialDependency} at three reduced times from $t^* = 0.09$ to 0.79. The radial velocity of the flow induced by the motion of the disk from the bottom wall is negative (thus inward) under the disk, zero at $r = 0$, maximum in absolute value near the edge of the disk at $r \simeq D/2$, before vanishing outside the gap. At $t^*=0.09$, the velocity obtained experimentally is found weaker (in absolute value) than the velocity obtained from the numerical simulation. As in Sec. III, this discrepancy is due to the shadow between the disk and the wall, which limits the accuracy of the PIV fields obtained when the disk is very close to the wall. The agreement between the experiment and the numerical simulation is recovered as soon as the disk has moved far enough away from the wall.  In the following, we consider the evolution of the radial velocity at the edge of the disk $r = D/2$. 

Examples of the vertical distribution of the radial velocity $u_r(D/2,t^*)$ at the edge of the disk are shown in figure \ref{fig:FigXX_HorizontalVelocityDPB} at six reduced times from $t^* = 0.1$ to 0.8. As the disk moves away from the wall, the flow between the disk and the wall corresponds to the superimposition of a radial flow entering by suction and a flow induced by the vortex ring in formation. This is particularly apparent at $t^* = 0.5$ and 0.8. We recall that, for a disk moving toward the wall, the radial velocity between the disk and the wall is always positive (see figure \ref{fig:FigXX_HorizontalVelocityDPH}). In the present case, where the disk is moving away from the wall, the radial velocity is negative near the bottom wall where the fluid enters below the disk but positive just below the disk. This region of positive value radial velocity is caused by the circumferential winding velocity of the vortex that forms in the near wake of the disk.

 \begin{figure}[t!]
     \centering
     \includegraphics[width=\linewidth]{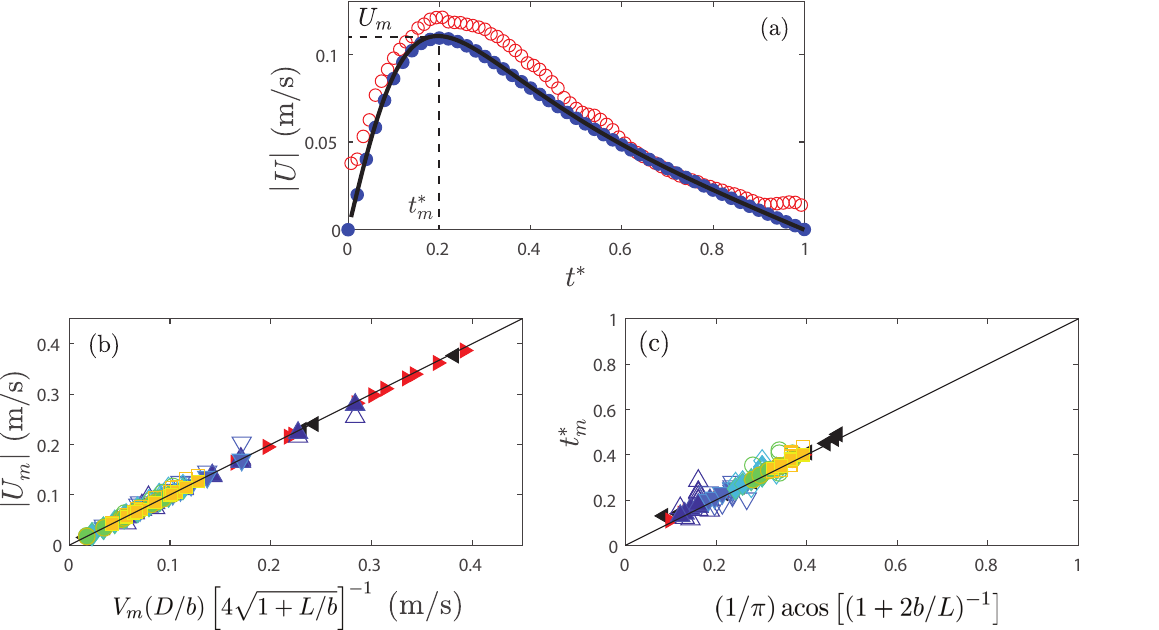}
     \caption{(a) Typical time evolution of the mean radial velocity $U$ at $r = D/2$ between the disk and the wall for the case of Fig. 12. Open symbols correspond to experimental measurements, filled symbols correspond to numerical simulations, and the solid line corresponds to the model in Eq. (\ref{eq:radial_velocity_away}). (b) Measured maximum radial velocity $U_m$ at $r = D/2$ from experiments (open symbols) and simulations (filled symbols) as a function of the predicted value from Eq. (\ref{eq:maximum_radial_velocity_away}). (c) Measured dimensionless time for the maximum radial velocity as a function of the predicted dimensionless time from Eq. (\ref{eq:maximum_time_away}).}
     \label{fig:FigXX_Away_UMeanAndTimeMaxDPB}
\end{figure}

An example of the time evolution of the absolute value of the radial velocity $U(D/2,t^*)$ averaged over the gap width $h(t^*)$ at the edge of the disk is shown in figure \ref{fig:FigXX_Away_UMeanAndTimeMaxDPB}(a) for the experimental measurements and the numerical simulation, together with the prediction 

 \begin{equation}
     U(r,t^*) = \frac{-r V_m \sin (\pi t^*)}{2 b + L[1-\cos (\pi t^*)]},
     \label{eq:radial_velocity_away}
 \end{equation}
\noindent which can be obtained from Eq. (\ref{eq:general_radial_velocity}), and using Eqs (\ref{eq:DiskPosition})-(\ref{eq:DiskVelocity}) for the disk motion away from the wall. The simulation and the model are in excellent agreement and rather close to the experiments. We observe that $U(D/2,t^*)$ first increases and reaches a maximum value of $|U_m| \simeq 0.12~$ m/s during the first part of the motion, here at $t_m^* \simeq 0.2$, before decreasing. All values of the maximum radial velocity $|U_m|$ obtained experimentally and numerically for all the configurations investigated here are plotted in figure \ref{fig:FigXX_Away_UMeanAndTimeMaxDPB}(b) against the predicted values given by

\begin{equation}
     U_m = -\frac{D}{b}\frac{V_m}{4 \sqrt{1+L/b}}.
     \label{eq:maximum_radial_velocity_away}
\end{equation}

Despite the presence of the start-up vortex under the disk, the value given by Eq. (\ref{eq:maximum_radial_velocity_away}) is in good agreement with both the experimental measurements and numerical results. Since the $z$-distribution of the radial velocity under the disk displays both positive and negative values, the inflow must be larger to compensate for the outflow due to the presence of the vortex. The velocities in the space between the bottom wall and the vortex are, therefore, much larger than when the disk is approaching the wall. Finally, all values of the maximum time $t^{*}_m$ obtained from experimental measurements and numerical simulations for all configurations considered are shown in figure \ref{fig:FigXX_Away_UMeanAndTimeMaxDPB}(c) against the predicted maximum time 

\begin{equation}
     t^*_m = \frac{1}{\pi} \, \text{acos} \left(\frac{1}{1+2b/L}\right).
     \label{eq:maximum_time_away}
\end{equation}
Since $1+2b/L > 1$, $t^*_m$ is theoretically always smaller than 1/2, meaning that the maximum radial velocity occurs during the first half of the disk motion ($0 < t^* < 1/2$). The results reported in Fig. 13(c) indeed show that this is observed for all flow configurations considered in the experiments and the numerical simulations. A close inspection of all the data points reveals that the values from the experimental measurements agree a bit less with the predicted values than the values extracted from the numerical simulations. The small difference is likely due to the measurement error associated both with the shadow zone between the disk and the wall, and the low number of reliable velocity vectors obtained on the PIV fields when the disk is very close to the wall.

\subsection{Features of the starting vortex}

\begin{figure}[t!]
    \centering
    \includegraphics[width=\linewidth]{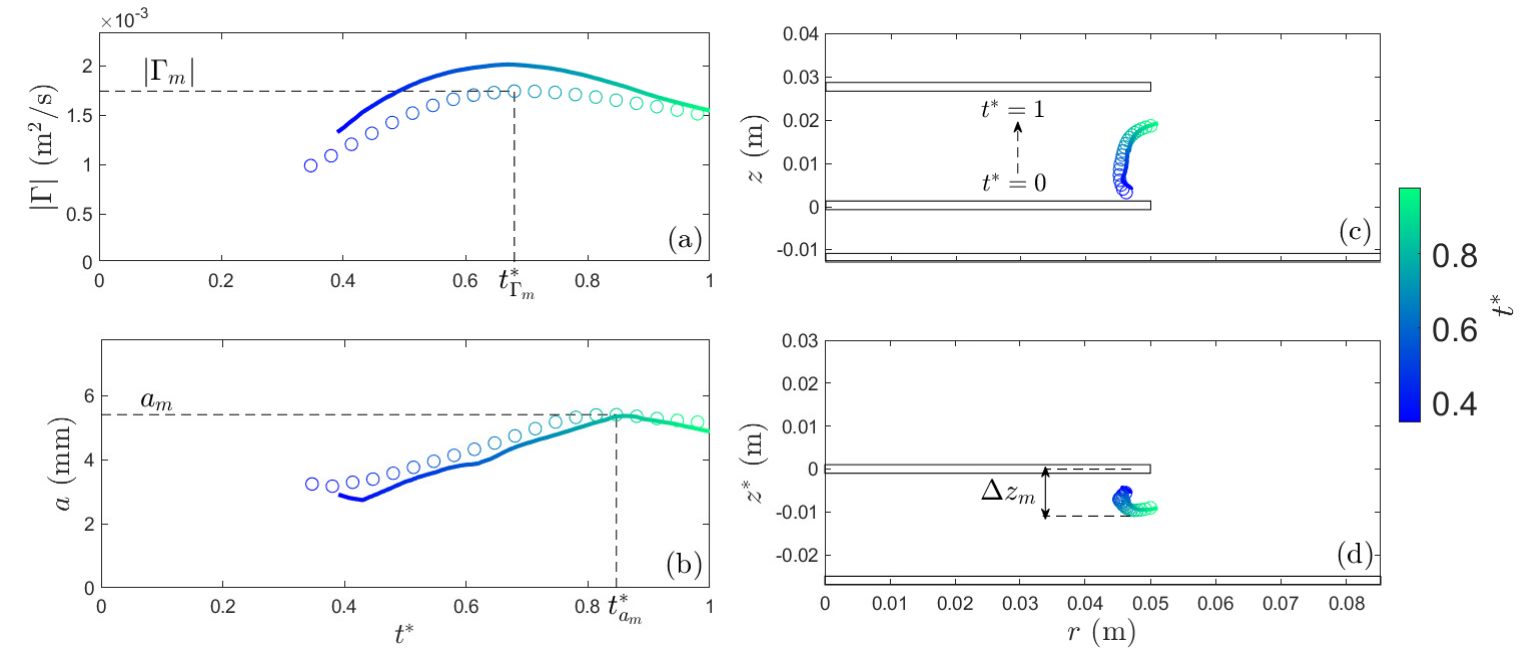}
    \caption{Time evolution of (a) the circulation $|\Gamma|$, (b) the radius $a$, and (c,d) the position ($r_G$, $z_G$) of the starting vortex in the reference frame of either (c) the laboratory or (d) the disk as it moves away from the wall for $L=2.8~$cm, $D = 10~$ cm, $\tau=2.5~$s, and $b=1~$cm ($Re = 1.8 \times 10^3$, $L/D = 0.28$, $b/D = 0.1$). The symbols (\protect\markerExpEvolution) represent experimental data, and the solid lines (\textcolor{blue}{\textbf{-}}) represent the corresponding numerical simulation data. The color bar indicates the dimensionless time $t^*$.}
    \label{fig:FigXX_VortexFeaturesDPBFlipped}
\end{figure}

\subsubsection{Time Dynamics}
A typical example of the time evolution of the properties of the starting vortex ring during its generation when the disk is moving away from the wall is shown in Fig. \ref{fig:FigXX_VortexFeaturesDPBFlipped} for the experiment and the corresponding numerical simulation with $L=2.8~$cm, $D = 10~$ cm, $\tau=2.5~$s, and $b=1~$cm ($Re = 1.8 \times 10^3$, $L/D = 0.28$, $b/D = 0.1$). We first note that there is a relatively good quantitative agreement between the features of the vortex in the simulation and the experiment, despite a difference of about 20\% for the circulation. In this example, the circulation and the core radius of the vortex ring increase with time and reach their maximum values $|\Gamma_m|\simeq1.7\times10^{-3}~$m$^2$/s and $a_m\simeq5.4~$mm at $t^*_\Gamma \simeq$ 0.68 and $t^*_a \simeq 0.85$, respectively, before decreasing. Both times $t^*_\Gamma$ and $t^*_a$ are here very different from the time $t^*_m$ at which the inward velocity in the gap is maximum. Figures \ref{fig:FigXX_VortexFeaturesDPBFlipped}(c)-(d) show the trajectory of the vortex: it moves first essentially vertically keeping close to the edge of the disk and then begin to moves outwards due to the deceleration of the disk.

\begin{figure}[t!]
    \centering
    \includegraphics[width=0.9\linewidth]{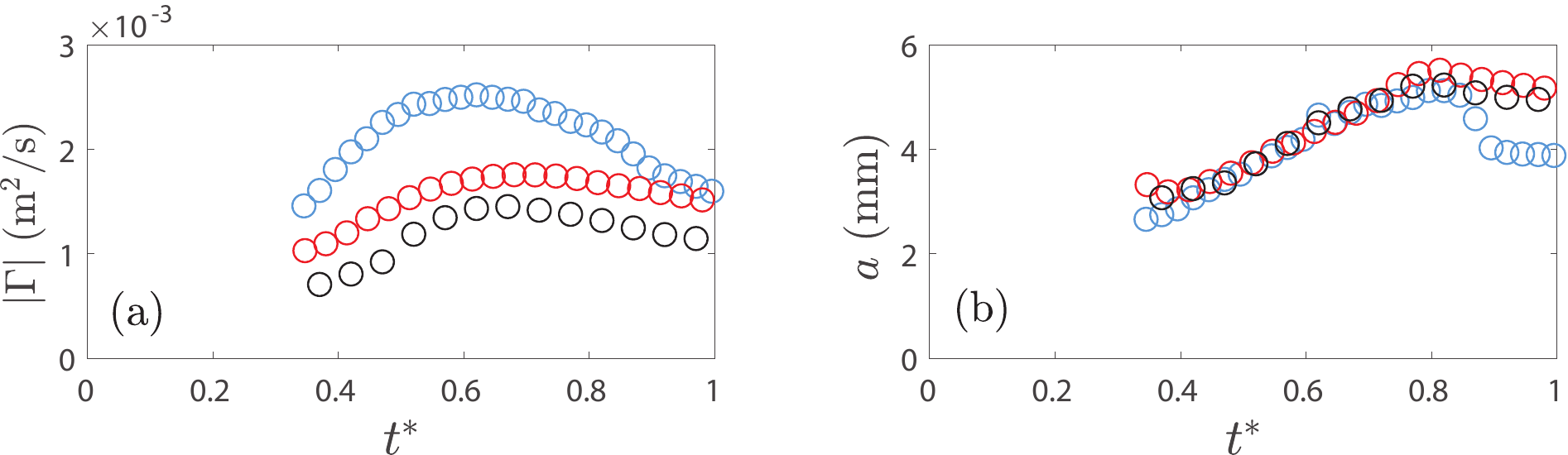}
    \caption{Time evolution of (a) the circulation $|\Gamma|$ and (b) the core radius $a$ of the starting vortex measured experimentally for a disk moving away from a wall for $D=10~$cm, $L=2.8~$cm and $\tau=2.5~$s ($Re = 1.8 \times 10^3$, $L/D = 0.28$), and for the starting distance $b=0.2$ cm $ = 0.02D$ (blue), $b = 1$ cm $ = 0.1D$ (red), and $b = 20$ cm $ = 2D$ (black).}
    \label{fig:Fig9_Away_VortexFeatures_DifferentHeights}
\end{figure}

To check the effect of the proximity of the wall, the time evolution of the circulation and the radius of the starting vortex for three initial distances $b$ with all other parameters kept constant is shown in Figs. \ref{fig:Fig9_Away_VortexFeatures_DifferentHeights}(a) and \ref{fig:Fig9_Away_VortexFeatures_DifferentHeights}(b), respectively. Similarly to the observation made in section III for a disk moving toward a wall, the maximum circulation also increases here when $b$ decreases. In contrast, the core radius does not seem to be significantly affected by the proximity of the wall.

\subsubsection{Scaling laws}
 \begin{figure}[t!]
     \centering
     \includegraphics[width=0.9\linewidth]{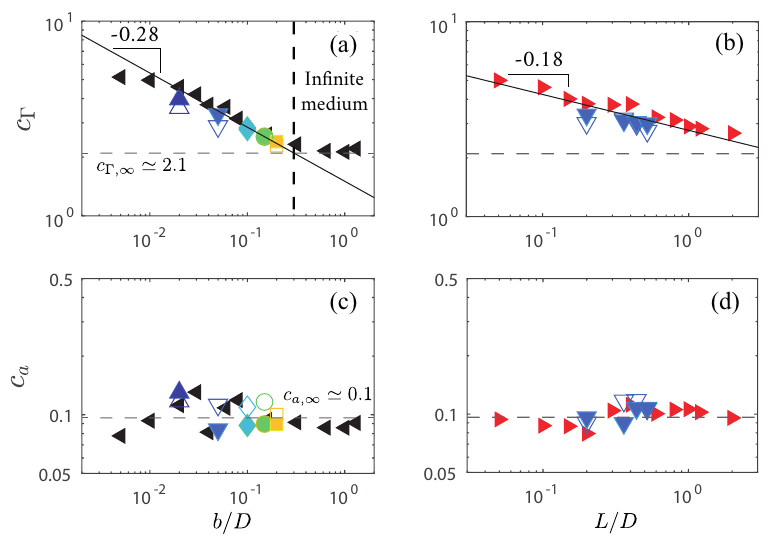}
     \caption{(a) Dimensionless maximum circulation $c_\Gamma$, (c) dimensionless maximum radius $c_a$ as a function of $b/D$ when $L/D = 0.28$, and (b) $c_\Gamma$ and (d) $c_a$ as a function of $L/D$ when $b/D = 0.05$ for a disk moving away from a wall. The horizontal dashed lines represent the values in the infinite medium: (a), (b) $c_{\Gamma, \infty} \simeq 2.1$, (c), (d) $c_{a, \infty} \simeq 0.1$. The solid lines are given by the equations: (a) $c_\Gamma = 1.5 \, (b/D)^{-0.28}$ and (b) $c_\Gamma = 2.8 \, (L/D)^{-0.18}$.}
     \label{fig:Fig10_Away_cAll}
\end{figure}

In this section, we focus on the maximum circulation and core radius of the starting vortex generated in the wake of the disk moving away from the wall. For this purpose, some numerical simulations have been performed by varying $b/D$ at constant $L/D=0.28$ and by varying $L/D$ at constant $b/D=0.05$, as shown in the sixth and seventh rows of Table \ref{tab:Symbols}.

In Fig. \ref{fig:Fig10_Away_cAll}(a), we observe that $c_\Gamma$ is constant and equal to the value that prevails in an infinite medium for $b/D \gtrsim 0.3$ but increases with decreasing $b/D$ for $b/D \lesssim 0.3$. In that range, a power-law fit $c_\Gamma \propto (b/D)^{-0.28}$ captures the data reasonably well over about a decade of $b/D$. Note that this scaling is the same as the one already observed when the disk translates toward the wall, with almost the same prefactor.
In Fig. \ref{fig:Fig10_Away_cAll}(b) we observe that $c_\Gamma$ decreases with increasing $L/D$, so that $c_\Gamma~\sim~(L/D)^{-0.18}$.  For larger $L/D$, the disk evolves over a longer distance without feeling the influence of the wall, which explains why $c_\Gamma$ is a decreasing function of $L/D$.

In contrast to the case of a disk moving toward a wall, the coefficient $c_a$ does not show any significant dependence with either $b/D$ or $L/D$. We observe in Fig. \ref{fig:Fig10_Away_cAll}(c) and (d) that all data collapse around the unbounded value $c_a \simeq 0.1$ with a deviation of $\pm 30\%$. The radius of the vortex does indeed not seem to be limited by the wall since the distance between the disk and the wall increases faster than the vortex radius. In short, the wall does not prevent the vortex from developing. 

In summary, the circulation of the vortex generated by a disk moving away from the wall is larger than in an infinite medium, while the core radius is unaffected. In such a case, the non-dimensional coefficient $c_\Gamma$ is larger than the value of 2.1 observed in an infinite medium and follows the scaling $c_\Gamma \sim (b/D)^{-0.28}(L/D)^{-0.18}$ in the range of parameters considered here where we have varied $b/D$ at $L/D$ = 0.28 and $L/D$ at $b/D$ = 0.05.  

 \begin{figure}[t!]
     \centering
     \includegraphics[width=0.6\linewidth]{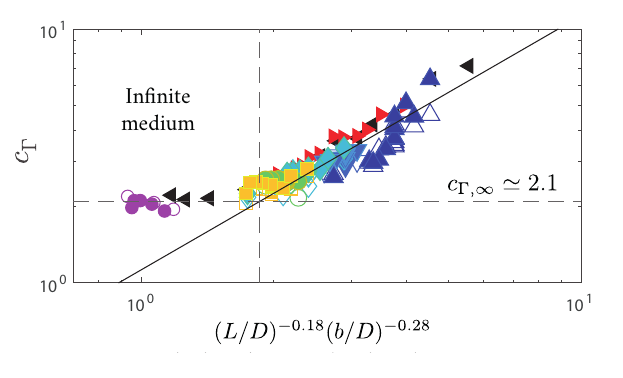}
     \caption{Dimensionless maximum circulation $c_\Gamma$ for a disk moving away from a wall as a function of $(L/D)^{-0.18}(b/D)^{-0.28}$. The dashed line indicates its value in the infinite domain $c_\Gamma \simeq 2.1$, and the equation of the solid line is $c_\Gamma = 1.1 (L/D)^{-0.18}(b/D)^{-0.28}$. All open and filled symbols correspond to the experiments and simulations reported in Table I, except the symbols (\protect\markercGammaInfinityFull) corresponding to the experimental and numerical results obtained by \cite{Steiner_2023} in an ``infinite'' medium for $b \simeq 20~$cm.}
     \label{fig:Fig11_Away_cGammaMax}
\end{figure}

Similarly to the previous section, we test these empirical scalings for all the configurations of different $L$, $D$, and $b$ reported in Table I, by plotting $c_\Gamma$ in Fig. \ref{fig:Fig11_Away_cGammaMax} as a function of $(b/D)^{-0.28}(L/D)^{-0.18}$. All experimental and numerical data collapse rather well on a master curve that shows a plateau corresponding to the unbounded limit case followed by an increasing trend so that the evolution of the coefficient $c_\Gamma$ is given by

\begin{equation}
c_\Gamma = \left\{ 
\begin{array}{l}
2.1 \\
\\
1.13 \, (L/D)^{-0.18}(b/D)^{-0.28}
\end{array}
\quad \mathrm{for} \quad (L/D)^{-0.18}(b/D)^{-0.28} \,\left\{
\begin{array}{l}
< 1.9 \\
\\
> 1.9
\end{array}
\right.
\right.
\label{eq:adim_circulation_away}
\end{equation}

The maximum circulation of the starting vortex in the near-wall configuration can be rewritten in dimensional form as

\begin{equation}
    |\Gamma_m| = 1.1 \, L^{1.15}D^{1.12}b^{-0.28}\tau^{-1}.
\end{equation}

Figure 17 shows that the maximum circulation of the vortex ring is larger when the disk is translating away from a wall than when it is translating in an infinite medium. In addition, $\Gamma_m$ scales differently with the setup parameters. The effect of the disk stroke $L$ on the circulation is less pronounced with a much weaker scaling exponent (1.15 instead of 4/3), and the effect of the disk diameter $D$ is much more pronounced with a larger scaling exponent (1.12 instead of 2/3). The scaling of the circulation for a disk translating away from a wall is also quite different from the scaling for a disk translating toward a wall. Indeed, the effect of $L$ is slightly smaller with a smaller scaling exponent (1.15 instead of 1.23), and the effect of $D$ is slightly larger with a larger scaling exponent (1.12 instead of 1.05), so the effect of $b$ is the same with the same scaling exponent ($-0.28$). The different scalings of the maximum circulation $\Gamma_m$ and radius $a_m$ of the vortex are summarized in Table II for a disk motion in an unbounded fluid and toward or away from a wall. We believe that these scaling laws are not specific to the present sinusoidal motion of the disk but the numerical prefactor would be slightly different for another type of motion such as with a uniform velocity for instance.

\begin{table}[h]
\centering
\setlength{\tabcolsep}{6pt}
\begin{tabular}{ |c c c c| } 
 \hline
  & unbounded & toward wall & away from wall \\
 \hline
$|\Gamma_m|$ & $2.1 \, L^{4/3}D^{2/3}\tau^{-1}$ & $1.3 \, L^{1.23}D^{1.05}b^{-0.28}\tau^{-1}$ & $1.1 \, L^{1.15}D^{1.12}b^{-0.28}\tau^{-1}$ \\
$a_m$ & $0.1 \, L^{2/3}D^{1/3}$ & $0.07 \, L^{0.55}D^{0.59}b^{-0.14}$ & $0.1 \, L^{2/3}D^{1/3}$ \\
 \hline
\end{tabular}
 \caption{Scaling laws for the maximum circulation $\Gamma_m$ and radius $a_m$ of the starting vortex for a disk motion in an unbounded fluid, and toward or away from a wall.}
 \label{tab:ScalingLaws}
\end{table}

\section{Conclusion}
\label{SecV}
In this paper, we have investigated the influence of a solid wall on the properties of the vortex ring generated in the wake of a disk translating toward and away from a wall. Axisymmetric numerical simulations show good agreement with PIV results during all the disk motion and after it stops up to $t^*\simeq2$. This suggests that, in the range of parameters considered here where $Re \lesssim 10^4$ and $L/D \ll 4$, non-axisymmetric fluctuations in the flow are not dominant in the generation of a vortex ring by the disk translation.
When the disk moves toward the wall, the decreasing distance between the disk and the wall results in an enhanced radial flow so that the vortex ring generated in the wake of the disk has a larger circulation and radius compared to the unbounded case where the disk is far from any boundary. The flow enters the vortex at a faster rate, causing its radius and circulation to grow faster. After the disk stops, the vortex moves radially and impacts the wall. Stopping vortex rings are generated and rotate around the starting vortex, leading to a modification of its trajectory, similar to the behavior in an unbounded medium.

In the opposite situation, where the disk moves away from the wall, we observe that the presence of the wall increases the circulation of the vortex ring, but the radius of the vortex core remains unchanged. This result means that the same amount of fluid enters the vortex as in the infinite medium configuration. As the circulation increases, the circumferential speed of the vortex also increases due to the interaction of the vortex with the wall. In addition, the large velocity flow between the wall and the vortex contributes to the growth of the circulation of the vortex.

In conclusion, this work sheds light on the influence of a wall on the properties of the vortex ring generated in the wake of a translating disk. These findings contribute to a better understanding of fluid-solid interaction phenomena and their implications in various applications such as locomotion. Scaling laws governing the maximum circulation and core size of the starting vortex have been established. They may be used, for instance, to characterize the deformation and erosion of a sediment bed deposited at the bottom wall as an interesting prospect to better understand the behavior of flatfish hiding in sand. In that case, the collision of the vortex ring with the bottom may give rise to flow instabilities that lead to non axisymmetric erosion patterns as already shown in \cite{Eames2000, Munro2009, Masuda2012, Yoshida2012, Yoshida2014, Yoshida2015}.

\section*{Acknowledgement}
The authors thank Johannes Amarni, Alban Aubertin, Lionel Auffray, and Rafaël Pidoux for their work on the experimental setup and L. Talon for his financial support.

\bibliography{bibliography}

\end{document}